%% file: main.tex
\documentclass[journal]{IEEEtran}

\usepackage{todonotes}
\usepackage{amssymb,amsfonts}
\usepackage{mathtools}
\usepackage{algorithm}
\usepackage{algpseudocode}
\usepackage{array}
\usepackage{graphicx}
\usepackage{xspace}
\usepackage[nopostdot,acronym,shortcuts,nonumberlist]{glossaries}
\usepackage{booktabs}
\usepackage[hyphens]{url}
\usepackage{enumitem}
\usepackage{cite}
\usepackage{comment}
\usepackage[hidelinks, breaklinks]{hyperref}
\usepackage{subcaption}
\usepackage{setspace}
\usepackage[table, dvipsnames]{xcolor}
\usepackage{makecell}

\definecolor{customblue}{HTML}{1f78b4}
\definecolor{customorange}{HTML}{E87722}
\definecolor{customgreen}{HTML}{658D1B}
\definecolor{custompurple}{HTML}{772583}
\definecolor{customred}{HTML}{E6001A}
\definecolor{customgray}{HTML}{4B4B4B}
\definecolor{custombeige}{HTML}{B8B184}

\newcommand{\ao}[1]{\textcolor{black}{#1}}
\newcommand{\wds}[1]{\textcolor{black}{#1}}
\definecolor{tudablue}{HTML}{4d80c2}
\newcommand{\revision} [1]{{\textcolor{black}{{#1}}}}

\DeclareMathOperator*{\argmin}{arg\,min}

\def\sysname{\textsc{SkyLink}\xspace}
\renewcommand{\refeq}[1]{(\ref{#1})}
\input{resources/B_acronyms}

\begin{document}

\pagestyle{empty}

\vspace*{3.5cm}
\hspace{0.2cm}
\fboxrule1.5pt
\framebox[17cm]{
\parbox{16cm}{
\vspace{0.3cm}

\copyright 2025 IEEE.  Personal use of this material is permitted.  Permission from IEEE must be obtained for all other uses, in any current or future media, including reprinting/republishing this material for advertising or promotional purposes, creating new collective works, for resale or redistribution to servers or lists, or reuse of any copyrighted component of this work in other works.

\vspace*{0.3cm}
}}

\title{\sysname: Scalable and Resilient Link Management in \gls{leo} Satellite Networks}

\author{Wanja de Sombre \IEEEauthorrefmark{1}, Arash Asadi \IEEEauthorrefmark{2}, Debopam Bhattacherjee \IEEEauthorrefmark{3}, Deepak Vasisht \IEEEauthorrefmark{4}, Andrea Ortiz \IEEEauthorrefmark{5}
\IEEEauthorblockA{
\IEEEauthorrefmark{1}Communications Engineering Lab, Technical University of Darmstadt, Germany, w.sombre@nt.tu-darmstadt.de\\
\IEEEauthorrefmark{2}WISE Lab, Delft University of Technology, The Netherlands, a.asadi@tudelft.nl\\
\IEEEauthorrefmark{3}Microsoft Research, India, debopamb@microsoft.com\\
\IEEEauthorrefmark{4}University of Illinois Urbana-Champaign, USA, deepakv@illinois.edu\\
\IEEEauthorrefmark{5}Institute of Telecommunications, Vienna University of Technology, Austria, andrea.ortiz@tuwien.ac.at}
\thanks{This work was funded by the BMBF project Open6GHub under grant 16KISKO14, by DAAD with funds from the German Federal Ministry of Research, Technology and Space BMFTR, and by the LOEWE Center emergenCity under grant LOEWE/1/12/519/03/05.001(0016)/72. The work of Andrea Ortiz was funded by the Vienna Science and Technology Fund WWTF Grant ID 10.47379/VRG23002.}}

\maketitle

\input{sections/0_abstract}
\input{sections/1_introduction}
\input{sections/2_sota}
\input{sections/3_system_model}

\input{sections/4_problem}
\input{sections/5_solution}
\input{sections/6_simulator}
\input{sections/7_results}
\input{sections/8_conclusion}

\bibliographystyle{IEEEtran}
\bibliography{IEEEabrv, resources/biblio}

\end{document}

%% file: resources/B_acronyms.tex
\newacronym{isl}{ISL}{Inter-Satellite Link}
\newacronym{gsl}{GSL}{Ground-Station-Satellite Link}
\newacronym{mab}{MAB}{Multi-Armed Bandit}
\newacronym{ucb}{UCB}{Upper Confidence Bound}
\newacronym{snr}{SNR}{signal-to-noise ratio}
\newacronym{tle}{TLE}{Two-Line Element}
\newacronym{fifo}{FIFO}{First-In-First-Out}
\newacronym{leo}{LEO}{Low Earth Orbit}
\newacronym{meo}{MEO}{Medium Earth Orbit}
\newacronym{rl}{RL}{Reinforcement Learning}
\newacronym{distUCB}{DistUCB}{Distance Upper Confidence Bound}
\newacronym{dqn}{DQN}{Deep Q-Network}

%% file: sections/0_abstract.tex
\begin{abstract}

The rapid growth of space-based services has established Low Earth Orbit (LEO) satellite networks as a promising option for global broadband connectivity. 
Next-generation LEO networks leverage inter-satellite links (ISLs) to provide faster and more reliable communications compared to traditional bent-pipe architectures, even in remote regions.
However, the high mobility of satellites, dynamic traffic patterns, and potential link failures pose significant challenges for efficient and resilient routing.
To address these challenges, we model the LEO satellite network as a time-varying graph comprising a constellation of satellites and ground stations. 
Our objective is to minimize a weighted sum of average delay and packet drop rate. 
Each satellite independently decides how to distribute its incoming traffic to neighboring nodes in real time. 
Given the infeasibility of finding optimal solutions at scale, due to the exponential growth of routing options and uncertainties in link capacities, we propose \sysname, a novel fully distributed learning strategy for link management in LEO satellite networks.
\sysname enables each satellite to adapt to the time-varying network conditions, ensuring real-time responsiveness, scalability to millions of users, and resilience to network failures, while maintaining low communication overhead and computational complexity.
To support the evaluation of \sysname at global scale, we develop a new simulator for large-scale LEO satellite networks.
For 25.4 million users, \sysname reduces the weighted sum of average delay and drop rate by $29\%$ compared to the bent-pipe approach, and by $92\%$ compared to Dijkstra.
It lowers drop rates by $95\%$ relative to $k$-shortest paths, $99\%$ relative to Dijkstra, and $74\%$ compared to the bent-pipe baseline, while achieving up to $46\%$ higher throughput.
At the same time, \sysname maintains constant computational complexity with respect to constellation size.

% Extensive experiments show that \sysname significantly outperforms traditional routing strategies, reducing packet drop rates by at least $74\%$ and improving throughput by up to $46\%$, while maintaining constant computational complexity with respect to network size. Our results demonstrate \sysname's potential to enable efficient, resilient, and scalable broadband communication over next-generation satellite networks.
\end{abstract}

%% file: sections/1_introduction.tex
\section{Introduction}
\label{sec:intro}

% General Introduction Space Broadband
\IEEEPARstart{D}{riven} by lower satellite development and deployment costs, this century has seen the rapid growth of space-based services. As a result, space broadband services have become widely accessible \cite{costred1, costred2, costred3, costred4, costred5}. In fact, multiple reports highlight this sector as a key growth area with a strong cumulative growth rate \cite{finreport_1, finreport_2, finreport_3, finreport_4}.
% Bent-Pipe vs. ISLs
Early satellite broadband services used bent-pipe connectivity, i.e.,  data traveled from a user terminal to a satellite and directly back to a ground station \cite{Ma2023}. However, this method becomes impractical in remote areas without nearby ground stations. Consequently, providers have started to shift towards \glspl{isl}, where data is relayed between satellites at the speed of light before reaching the ground~\cite{starlink_lasers, kuiper_lasers}. 
Advantages of using \glspl{isl} include reduced delay and higher bandwidth, which is essential for broadband networks~\cite{optical_isls}. %Moreover, 

\input{sections/2a_sota_table}

%Effective \gls{isl} use demands dynamic link management which is challenging due to rapidly changing topologies. Consequently, modern \gls{leo} satellite networks rely on regenerative payloads\footnote{Enabling them with on-board signal processing and routing.} allowing for active data flows management and real-time routing (cf. 3GPP TR 38.811, \cite{3gpp38811}). However, finding globally optimal link management solutions is a complex problem. \gls{leo} satellite networks are highly mobile, with satellites orbiting at speeds of up to $26,000$~km/h at an altitude of $1,200$~km. Additionally, the number of routing paths grows exponentially with the network size. Therefore, the fact that \gls{leo} satellite networks often consist of several hundreds of satellites significantly increases the complexity of deciding how to use the available links. This is compounded by the need to handle extremely high data rates, with traffic volumes reaching Terabits per second. Moreover, satellites must constantly react to changes in the network and failures which can occur due to system malfunctions or external factors such as solar storms~\cite{solar_storm, stablehierarchicalrouting}.

Efficient use of \glspl{isl} demands dynamic link management which is challenging due to continuously evolving topologies. Satellites therefore employ regenerative payloads\footnote{Enabling onboard signal processing and routing.}, which allow them to manage data flows and perform real-time routing (cf. 3GPP TR 38.811~\cite{3gpp38811}). Yet, identifying globally optimal link configurations remains difficult due to the exponential growth in routing options as the constellation scales to hundreds of satellites. This complexity is further increased by terabit-scale traffic demands and the need to continuously react to topology changes and failures, whether due to system faults or external events such as solar storms ~\cite{solar_storm, stablehierarchicalrouting}.

% Previous Approaches (Shortest Paths + DQN)
Link management in \gls{leo} satellite networks can be approached using classical shortest-path algorithms\cite{Zhang2022} or more recent machine-learning-based solutions~\cite{Roth2024, Zhou2024}. While shortest-path algorithms, such as Dijkstra’s or $k$-shortest path, are widely used in terrestrial networks, they fall short in LEO scenarios \cite{Zhang2022}. These algorithms often result in paths that overlap on the same \glspl{gsl}, creating bottlenecks in areas where large volumes of data need to be transmitted simultaneously or where the \glspl{gsl} have limited capacities. 
To avoid such bottlenecks, recent research has explored the use of \glspl{dqn} to redistribute traffic and ensure efficient network performance~\cite{Roth2024, Zhou2024}. 
However, both, shortest-path algorithms and learning solutions often rely on central controllers to monitor network failures and propagate updated routes to all nodes. 
The controllers are usually located on earth due to the limited computational capabilities of LEO satellites. 
As a result, continuous collection of global network state information is required causing slow response time to rapid topology changes, and introducing communication overhead. In highly dynamic networks, any delay in distributing updated control commands can result in inefficient link management, increased packet drop rate, and higher delays. 
Additionally, the dependency on a central controller reduces the resilience of the system, as failures need to be centrally detected and accounted for. 

A partial solution to the necessity for a central controller is to offload link management decisions to ground stations or higher-altitude \gls{meo} satellites, where computational resources are less constrained than on \gls{leo} satellites \cite{ma2024leo, cbo2023leo, lin2024large, esa2023ids}.
However, communication to ground stations or \gls{meo} satellites introduces additional delay \cite{Keller1998} and still requires real-time network state information, which is difficult to maintain given the high mobility and frequent handovers. Additionally, failures in connections to \gls{meo} satellites can lead to substantial network disruptions, as link management decisions for entire regions might become inaccessible. Instead of relying on a central controller, pre-trained models can be periodically distributed across the network. However, this comes at the cost of slower response times to disturbances.

% Our solution
The development of a fully distributed, failure-resilient, and scalable link management strategy is as a key research challenge for advancing \gls{leo} satellite networks.
Therefore, in this paper, we introduce \sysname, a distributed learning approach that directly tackles the scalability, resilience, and computational complexity of link management in \gls{leo} satellite networks. \sysname combines the \gls{mab} framework~\cite{Sutton2018} with tile coding to enable each satellite to autonomously and efficiently prioritize its communication links using only local information. By making real-time, decentralized decisions, \sysname minimizes the delay and  drop rate even under heavy traffic conditions and satellite failures,  and without relying on global coordination.

% Simulator
The evaluation of link management solutions for \gls{leo} satellite networks demands global scale analysis. However, existing simulations tools are often limited to network segments (see e.g. \cite{Roth2024, Deng2023}). Thus, to realistically benchmark \sysname{}, we introduce a novel simulator capable of modeling entire \gls{leo} satellite networks. Our simulator models global-scale traffic and allows us to validate \sysname's performance across various scenarios including millions of users, high traffic volumes, and network failures. 

% Contributions
In summary, the primary contributions of this work are:
\begin{itemize}
    \item We introduce \sysname{}, a fully distributed link management approach to jointly minimize the delay and drop rate in \gls{leo} satellite networks. Specifically, we break the complexity of the problem by using \gls{mab}-based learning including tile coding mechanisms. Due to its distributed nature, \sysname{} is scalable in network size and data volume and resilient to network failures.
    
    \item For a realistic evaluation, we present a new, extensive and fast simulator capable of  modeling \gls{leo} satellite networks globally. Our simulator includes advanced tools for space broadband simulations, detailed channel models for \glspl{isl} and \glspl{gsl}, a global data generation framework, a representation of network data streams, and diverse visualization options\footnote{\revision{The code is available under https://github.com/wanjads/SkyLinkSimulator.}}.
    
    \item We show via extensive evaluation that \sysname significantly reduces delay and drop rate in the \gls{leo} satellite network and increases the network's throughput compared to various reference schemes, even under scenarios of increased traffic load and satellite failures.
\end{itemize}

% Organisation
The rest of the paper is organized as follows. We discuss related work in Sec. \ref{sec:sota}. In Sec. \ref{sec:systModel} we introduce the system model and formulate the optimization problem in Sec. \ref{sec:problem}. \sysname is presented in Sec. \ref{sec:solution} and we describe the details of our simulator in Sec. \ref{sec:simulator}. The numerical evaluation is presented in Sec. \ref{sec:evaluation} and  Sec. \ref{sec:conclusion} concludes the paper.

%% file: sections/2a_sota_table.tex
\begin{table*}[t]
    \centering
    \scriptsize
    \renewcommand{\arraystretch}{2}
    \begin{tabular}{|c|*{18}{>{\centering\arraybackslash}m{0.35cm}|}}
        \hline
        & \makecell{\hspace{-0.1cm}\cite{Roth2024}} & \makecell{\hspace{-0.1cm}\cite{Zhou2024}} & \makecell{\hspace{-0.1cm}\cite{Wang2022}} & \makecell{\hspace{-0.1cm}\cite{Hou2022}} & \makecell{\hspace{-0.1cm}\cite{Zhang2023}} & \makecell{\hspace{-0.1cm}\cite{Gounder1999}} & \makecell{\hspace{-0.1cm}\cite{Huang2024}} & \makecell{\hspace{-0.1cm}\cite{Huang2024_2}} & \makecell{\hspace{-0.1cm}\cite{Deng2023}} & \makecell{\hspace{-0.1cm}\cite{Liu2020}} & \makecell{\hspace{-0.1cm}\cite{Soret2023}} & \makecell{\hspace{-0.1cm}\cite{Roth2023}} & \makecell{\hspace{-0.1cm}\cite{Handley2018}} & \makecell{\hspace{-0.1cm}\cite{Han2024}} & \makecell{\hspace{-0.1cm}\cite{Zhang2022}} & \makecell{\hspace{-0.1cm} \cite{stablehierarchicalrouting}} & \makecell{\hspace{-0.1cm} \cite{Lai2023}} & \cellcolor[gray]{0.9} \makecell{\hspace{-0.1cm} Ours} \\
        \hline
        \hline
        \makecell{Joint Minimization of \\ Delay \& Drop Rate} 
        & \checkmark & \checkmark & \checkmark & \checkmark & & & & \checkmark & \checkmark & \checkmark & \checkmark & \checkmark & & & \checkmark & \checkmark & \checkmark & \cellcolor[gray]{0.9}\checkmark \\
        \hline
        \makecell{Dynamic Network} 
        & & \checkmark & \checkmark & \checkmark & \checkmark & \checkmark & \checkmark & \checkmark & \checkmark & \checkmark & \checkmark & \checkmark & \checkmark & \checkmark & \checkmark & \checkmark & \checkmark & \cellcolor[gray]{0.9}\checkmark \\
        \hline
        \makecell{ISLs and GSLs} 
        & & & & \checkmark & & & \checkmark & \checkmark & \checkmark & \checkmark & \checkmark & \checkmark & \checkmark & \checkmark & \checkmark & \checkmark & \checkmark & \cellcolor[gray]{0.9}\checkmark \\
        \hline
        \makecell{Multi-Path Routing} 
        & & & & & \checkmark & \checkmark & \checkmark & \checkmark & \checkmark & \checkmark & \checkmark & \checkmark & \checkmark & \checkmark & \checkmark & \checkmark & \checkmark & \cellcolor[gray]{0.9}\checkmark \\
        \hline
        \makecell{Distributed Approach} 
        & \checkmark & & & & \checkmark & & & & \checkmark & \checkmark & \checkmark & & & & \checkmark & \checkmark & & \cellcolor[gray]{0.9}\checkmark \\
        \hline
        \makecell{Large Constellation} 
        & & & \checkmark & & & & & & & & & & \checkmark & \checkmark & \checkmark & \checkmark & \checkmark & \cellcolor[gray]{0.9}\checkmark \\
        \hline
        \makecell{Global Broadband Communication} 
        & & & & & & & & & & & & \checkmark & & & & & \checkmark & \cellcolor[gray]{0.9}\checkmark \\
        \hline
        \makecell{Resilience to Network Failures} 
        & & & & & & & & & & & & & & & \checkmark & \checkmark & \checkmark & \cellcolor[gray]{0.9}\checkmark \\
        \hline
    \end{tabular}
    \vspace{0.2cm}
    \caption{Summary of the related work on routing in \gls{leo} constellations.}
    \label{tab:sota}
\end{table*}

%% file: sections/2_sota.tex
\section{Related Work}
\label{sec:sota}
In this section, we review the related work on link management for \gls{leo} satellite networks. Table \ref{tab:sota} summarizes recent advancements in the field. 
The pioneering work of Gounder et al. introduced a $k$-shortest path algorithm for satellite networks \cite{Gounder1999}. Over the past five years, the increasing availability and deployment of large \gls{leo} satellite networks have sparked considerable interest within the research community. 

Contemporary studies already consider dynamic satellite networks using both \glspl{isl} and \glspl{gsl},
and allowing multi-path routing to distribute traffic \cite{Huang2024, Huang2024_2, Handley2018, Han2024, Deng2023, Liu2020, Soret2023, Zhang2022, Roth2023, stablehierarchicalrouting, Lai2023}. In these works, link management solutions considering different optimization goals have been investigated. Specifically, the joint optimization of delay and packet drop rate is studied in \cite{Deng2023, Liu2020, Soret2023, Zhang2022, Roth2023, stablehierarchicalrouting, Lai2023}, delay minimization is considered \cite{Han2024, Handley2018}, forwarding overhead reduction is investigated in \cite{Zhang2023}, and \cite{Huang2024_2} maximizes the network efficiency.

%in with only a few studies focusing on single metrics like delay or throughput \cite{Gounder1999, Zhang2023, Huang2024, Han2024, Handley2018}.
Although some of these works consider only relatively small \gls{leo} constellations, such as Iridium \cite{Deng2023, Liu2020, Huang2024_2} or Kepler \cite{Soret2023},  works like \cite{Huang2024, Roth2023} address networks with a couple of hundreds of satellites while \cite{ Handley2018, Han2024, Zhang2022, stablehierarchicalrouting, Lai2023} investigate mega \gls{leo} constellations. Mega constellations, such as Starlink or Kuiper, consist of thousands of satellites and are designed to deliver global broadband communication to millions of users. However, despite their focus on mega constellations, many studies consider only on a subset of communication paths, typically optimizing pre-defined source-destination pairs rather than  network-wide traffic patterns \cite{Han2024, Handley2018, Zhang2022, stablehierarchicalrouting}.
Only the authors of \cite{Lai2023} consider global broadband traffic. In particular, they investigate the satellite network's resilience through a hierarchical model relying on nearest-neighbor searches for route selection. However, the proposed model relies on a centralized architecture which can lead to congestion under high traffic. In fact, a critical gap in the literature is the lack of approaches that simultaneously adopt a distributed framework and focus on resilience against network failures.

%in mega constellations. Distributed approaches, as seen in works like \cite{Roth2024, Zhang2023, stablehierarchicalrouting, Deng2023, Liu2020, Soret2023, Zhang2022}, offer scalability and reduced reliance on central controllers, but often lack mechanisms to handle dynamic failures effectively.
In summary, while impressive progress has been made in routing for \gls{leo} satellite networks, challenges remain in achieving scalable and resilient solutions for large constellations. 

%Current research often focuses on specific traffic paths or small constellations, with limited consideration of network-wide dynamics and resilience. Distributed approaches offer potential for scalability but often fall short in addressing dynamic failures effectively. With \sysname{}, we aim to fill this gap by proposing a fully distributed, scalable and resilient link management solution for global broadband communications using \gls{leo} satellite networks.% and by evaluating it in a realistic environment.

%% file: sections/3_system_model.tex
\section{System Model}
\label{sec:systModel}

\subsection{LEO Satellite Networks}
\begin{figure}[t]
    \centering
    \includegraphics[width=\columnwidth]{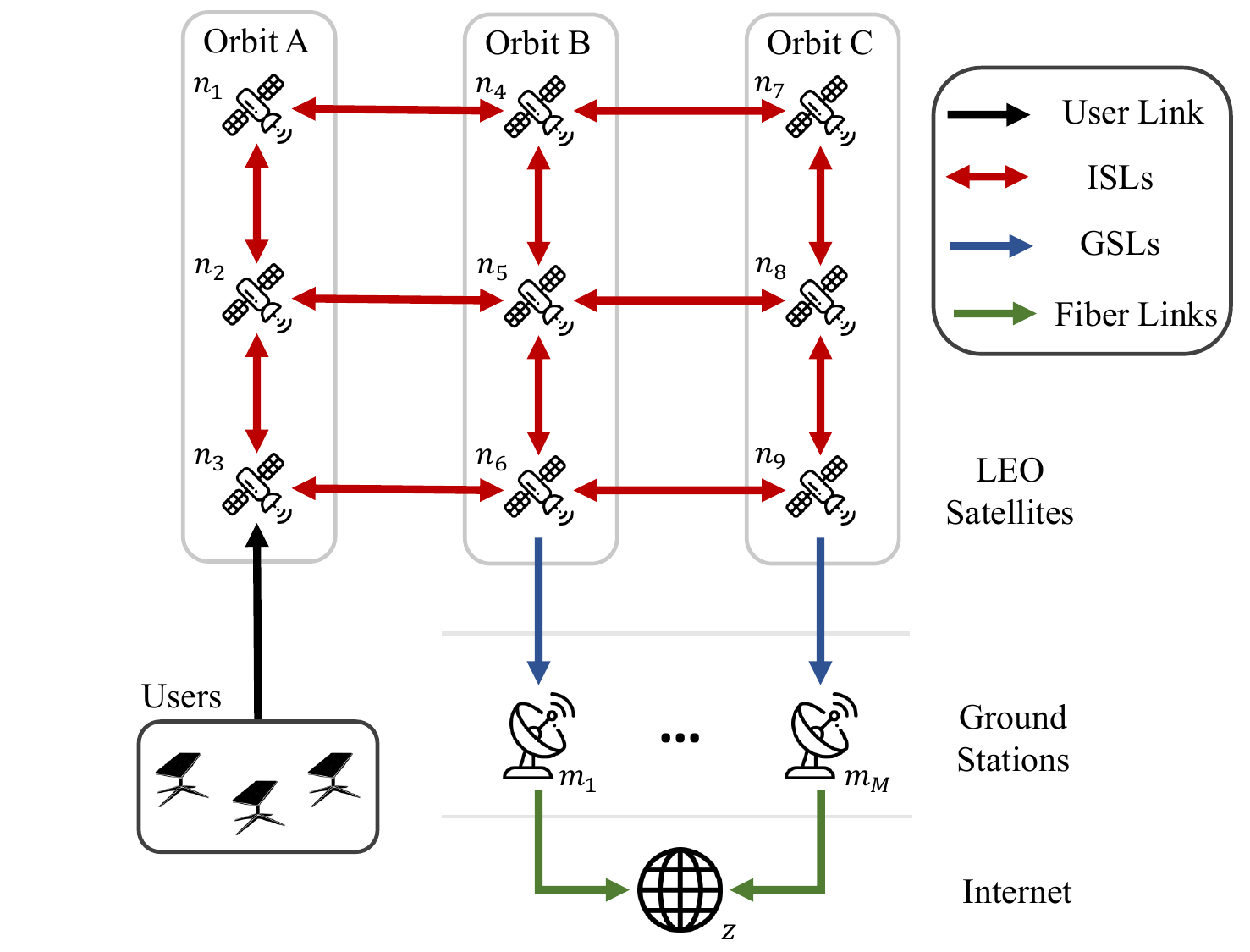}
    \caption{{Diagram of the considered scenario}}
    \label{fig:systemmodel}
\end{figure}

We focus on a communication scenario in which the satellites receive data from the users within their coverage area and route it to the internet. As illustrated in \autoref{fig:systemmodel}, we consider a \gls{leo} satellite network consisting of a set $\mathcal{N}$ of $N$ \gls{leo} satellites $\mathcal{N} = \{n_1, \ldots, n_N\}$ and a set $\mathcal{M}$ of $M$ ground stations $\mathcal{M} = \{m_1, \ldots, m_M\}$. The system operates in time slots, starting at $t=0$ and continuing until a finite time horizon $T$, with each slot lasting a fixed and constant duration $\tau$. 

Satellites establish both, full-duplex optical \glspl{isl} \cite{duplex} and half-duplex radio \glspl{gsl} and decide, 
in every time slot, which established \glspl{isl} and \glspl{gsl} to use to relay their incoming data. 
Considering current technology, we assume that each satellite establishes at most four \glspl{isl} to neighboring satellites \cite{SpaceX2018FCCFiling, Wang2023, Wu2024}, two of which are in the same orbit, while the remaining are neighboring satellites in adjacent orbits. Consequently, the \glspl{isl} form a $+$grid as indicated in \autoref{fig:systemmodel}. 
Due to irregularities in this grid, pending deployment or failures, the next neighboring satellite in a given direction might not be available for an \gls{isl}. 
In such cases, the satellite does not establish an \gls{isl} in that direction.
In every time slot, each ground station $m_i$ establishes \glspl{gsl} to the $\mu_i$ closest satellites, where $\mu_i$ is the number of antennas at ground station $m_i$. Satellites use these \glspl{gsl} to send data to the ground.
The ground stations establish connections to the internet via fiber optic links. 
Both, satellites and ground stations can simultaneously transmit data over their outgoing links and receive data from satellites. 

\subsection{Grid Model}
All the links among satellites and ground stations form a directed graph $\mathcal{G}_t=(\mathcal{V},\mathcal{E}_t)$ at each time slot $t$. 
Here, the vertex set $\mathcal{V}=\mathcal{N}\cup\mathcal{M}\cup\{z\}$ contains all satellites, ground stations, and the internet node $z$. 
The edge set $\mathcal{E}_t$ includes all links $(v, w)$ between any two nodes $v, w \in \mathcal{V}$ at time slot $t$. This includes \glspl{isl}, \glspl{gsl}, and the ground stations' links to the internet.
For any satellite or ground station $v$, the set $\mathcal{X}_{v,t}$ contains the paths to the internet. Each path $X$ is a sequence of links {$((w_0,w_1),(w_1,w_2), \dots ,(w_{n-1}, w_n))$. We call the target node of the last link in a path its \textit{terminal node}.} During time slot $t$, the data transmitted over each path $X$ is modeled as a continuous stream with a fixed rate $R_X$.
The variable $x_{(w,w'),t}$ represents the amount of data that satellite $w$ decides to transmit over the link $(w,w')$ in time slot $t$.
The vector $\mathbf{x}_t$ aggregates the decisions of all the satellites in time slot $t$ and is given by $\mathbf{x}_t = (x_{(v,w),t})_{(v,w) \in {\mathcal{E}_t}}$.
In time slot $t$, the set of outgoing links at node $v$ is denoted as $O_{v, t} = (\{v\} \times \mathcal{V}) \cap \mathcal{E}_t$.

\subsection{Data Generation Model}
\label{sec:dataGen}
We consider a stream-based model for data transmission, where each satellite receives individual data streams and relays them to the internet. 
The user data rates $R^\mathrm{g}_{n,t}$ received at satellite $n$ during time slot $t$ depend on the satellite's location and local time{, and are measured in bit/s}.
Higher rates occur over densely populated regions and peak hours, while lower rates are over remote areas and at night.
We divide the day into hourly intervals and the Earth's surface into a one-degree grid.
% For each time and each satellite, the data rate is calculated as the product of cell populations, average devices per person, and average data rate per person per second, based on the time of day. 
$R^\mathrm{g}_{n,t}$ is the sum of data streams of all the cells associated with satellite $n$ as
\begin{align}
\label{eq:userdata}
   R^\mathrm{g}_{n,t} = \mathrm{Pop}_{n,t} \cdot \mathrm{d} \cdot \nu \cdot \mathrm{t}_{n,t},
\end{align}
where in time slot $t$, $\mathrm{Pop}_{n,t}$ is the population count associated to satellite $n$, $\mathrm{d}$ is the average number of devices per person using the \gls{leo} network, $\nu$ is the average traffic per device and second and $\mathrm{t}_{n,t}$ is a factor scaling the traffic according to the local time of the day.

Satellites and ground stations have limited data buffers for storing data awaiting transmission, with each node $v$ having a capacity $Q^{\mathrm{max}}_v$.
If the outgoing data rate $R^\mathrm{out}_{v, t}$ of a node $v$ is less than the incoming data rate $R^\mathrm{in}_{v, t}$, the buffer at the receiving node fills up.
When the buffer is full, a uniform percentage of data from every incoming stream will be dropped to align the incoming stream size with the outgoing rate ($R^\mathrm{out}_{v, t}$ = $R^\mathrm{in}_{v, t}$). 
Our model focuses on the steady state of buffers, either filled or empty, disregarding the transitional phases of filling or emptying.
Incoming data at satellites without an outgoing link is dropped.

\subsection{Capacities and Delays}
Each link $(v, w) \in \mathcal{E}_t$ has a capacity $C_{(v,w), t}$.
If a satellite decides to transmit more data through a link than its capacity allows, i.e. if $x_{(v,w),t} > C_{(v,w), t}$, the excess amount of data is dropped.

The capacity $C_{(v,z), t}$ of the fiber optic link from a ground station $v$ to the internet $z$ is assumed to be constant.
\ao{To model the capacity $C_{(v,w), t}$ of an \gls{isl} between satellites $v,\, w \in \mathcal{N}$, we first calculate the received power $P_{\text{rx}, v, w, t}$ as the transmitted power $P_{\text{tx}}$ of satellite $v$ scaled by the pointing loss $L_\text{pointing}$ and the fraction of the transmitted beam intercepted by satellite $w$ as 
\begin{equation}
    {P_{\text{rx}, v, w, t} = P_{\text{tx}} \cdot L_{\text{pointing}} \cdot \frac{\pi \left( 0.5 \cdot \varnothing \right)^2}{\pi (d_{v, w, t} \cdot \theta)^2}}
\end{equation}
where $L_{\text{pointing}}$ results from the angular deviation between the transmitting and receiving antennas, $\varnothing$ is the aperture diameter of the receiver at satellite $w$, $d_{v, w, t}$ is the distance between the satellites, and $\theta$ is the beam divergence angle. Note that this model assumes a canonical beam and uniform beam intensity.}
The noise power is calculated as
\begin{equation}
    P_{\text{$\sigma$},v,w} = k \cdot T_{\sigma} \cdot B_\mathrm{ISL}
\end{equation}
where $k$ is the Boltzmann constant, $T_{\sigma}$ is the noise temperature, and $B_\mathrm{ISL}$ is the bandwidth of the link between $v$ and $w$.
$C_{(v,w), t}$ is then obtained using the Shannon-Hartley theorem:
\begin{equation}
\label{eq:capacity_isl}
    C_{(v,w),t} = \lambda \cdot B_{\text{ISL}} \cdot \log_2 \left(1 + \frac{P_{\text{rx}, v, w, t}}{P_{\text{$\sigma$},v,w}} \right),
\end{equation}
where, $\lambda < 1$ represents a scaling factor to account for the fact that only part of the link’s total capacity can be allocated for data transmission from the user to the internet.

The capacity $C_{(v,w), t}$ for the \glspl{gsl} between satellite $v$ and ground station $w$ is also modeled using the {Shannon-Hartley theorem}. In this case, the total received power $P_{\text{rx}}$ at the ground station is computed as 
\begin{equation} 
P_{\text{rx}} = \exp\left(\frac{\log(10) \cdot (\text{EIRP} - \text{FSPL} + G_{\text{rx}} - A_{\text{atmos}})}{10}\right), 
\end{equation}
where $\text{EIRP}$ is the equivalent isotropic radiated power transmitted by the satellite, $\text{FSPL}$ denotes the free-space path loss, $G_{\text{rx}}$ is the gain of the ground station's receiver antenna, and $A_{\text{atmos}}$ represents the atmospheric attenuation factor. The free-space path loss is given by

\begin{equation} 
\text{FSPL} = 20 \log_{10} \left(\frac{4 \pi d_{v,w, t} \cdot f_{\text{c}}}{c}\right), 
\end{equation}
where $f_{\text{c}}$ is the carrier frequency and $c$ is the speed of light. The atmospheric attenuation $A_{\text{atmos}}$ depends on the elevation angle between the satellite and the ground station, with lower elevation angles resulting in greater attenuation due to the longer path through the atmosphere.
The noise power $P_{\text{$\sigma$,v,w}}$ on the \glspl{gsl} is calculated as
\begin{equation} 
P_{\text{$\sigma$},v,w} = k \cdot T_{\text{sky}} \cdot B_{\text{GSL}}, 
\end{equation}
where $B_{\text{GSL}}$ is the bandwidth of the ground-satellite link, and $T_{\text{sky}}$ is the sky noise temperature. The sky noise temperature is influenced by atmospheric attenuation and is computed as
\begin{equation} 
%T_{\text{sky}} = T_{\text{mr}} \cdot \left(1 - 10^{-\frac{A_{\text{atmos}}}{10}}\right) + 2.7 \cdot 10^{-\frac{A_{\text{atmos}}}{10}}, 
T_{\text{sky}} = T_{\text{mr}} \cdot \left(1 - 10^{-{A_\text{atmos}}/{10}}\right) + 2.7 \cdot 10^{-{A_{\text{atmos}}}/{10}}, 
\end{equation}
where $T_{\text{mr}}$ represents the microwave background radiation temperature. %Finally, the capacity $C_{(v,w), t}$ of the \gls{gsl} is determined using the Shannon-Hartley theorem:
%\begin{equation}
% \label{eq:capacity_gsl} C_{(v,w),t} = B_{\text{GSL}} \cdot \log_2 \left(1 + \frac{P_{\text{rx}}}{P_{\text{$\sigma$}}} \right). 
% \end{equation}

\revision{Here, practical hardware effects can be incorporated by reducing the effective $\mathrm{EIRP}$ and/or $G_{\mathrm{rx}}$ and/or as an increase of the effective noise temperature. Examples for such effects can be found in the literature (see e.g. \cite{li2025hardware}). Hence, hardware non-idealities affect capacity through the resulting $\frac{P_{\mathrm{rx}}}{P_{\sigma}}$ and are implicitly reflected in the \glspl{gsl}' qualities.}

%% file: sections/4_problem.tex
\section{Problem Formulation}
\label{sec:problem}

% General Formulation
Our objective is to find a link management solution that minimizes the average cost given by a weighted sum of delay and drop rate. This formulation ensures that the network operates efficiently by avoiding trade-offs where reducing delay significantly increases data loss or minimizing drop rate causes excessive delays. To rigorously define the problem, we first introduce necessary definitions.

% Drop Rate
The drop rate $\zeta_{v,t}$ of a node $v$ in time slot $t$ is defined as
\begin{equation}
    \zeta_{v,t} = 1 - \tilde{R}_{v,t},
\end{equation}
where $\tilde{R}_{v,t}={R^\mathrm{s}_{v,t}}/{R^\mathrm{g}_{v,t}}$ is the share of traffic that was successfully routed to the internet.
$R^\mathrm{s}_{v,t}$ is the actual amount of user data traffic  successfully routed to the internet and ${R^\mathrm{g}_{v,t}}$ is defined in \refeq{eq:userdata}.
\revision{We do not consider single packets. Drop rates are computed directly at the rate level in our stream-based model. This modeling choice allows us to scale to hundreds of satellites and tens of millions of users while preserving the congestion, buffering, and delay phenomena that drive routing decisions.}
Data streams are dropped due to low link capacity, whenever they are transmitted in a loop, if they reach a node without outgoing links, or if their delay exceeds the maximum tolerable delay $T_\mathrm{max}$.

% Delay
The {propagation} delay $D^\mathrm{Tx}_{(v,w),t} \in \mathbb{R^+}$ for the link between $v \in \mathcal{N}$ and $w\in \mathcal{N}\cup \mathcal{M}$ is approximated as 
\begin{align}
    D^\mathrm{Tx}_{(v,w),t}& := \frac{d_{v,w, t}}{c}.
\end{align}
{For simplicity, the propagation delay $D^\mathrm{Tx}_{v,t}$ of the fiber optic link between ground station $v$ and the internet is modeled as a random variable drawn from a uniform distribution, with fluctuations introduced by Gaussian noise in each time slot. While the choice of ground station does influence the actual delay, this effect is beyond the scope of this work.}

The queuing delay $D^\mathrm{q}_{v,t}$ of node $v\in \mathcal{N}\cup \mathcal{M}$ is the average duration the data remains in the node's data buffer before being forwarded. 
Assuming a First-In-First-Out queue, $D^\mathrm{q}_{v,t}$ is:
\begin{equation}
    D^\mathrm{q}_{v,t} := 
    \begin{cases}
        0  & \text{ if } \Delta_C \leq 0 \\
        \frac{Q^{\mathrm{max}}_v}{\sum_{(v,w) \in O_{v,t}}\min(x_{(v,w),t}, C_{(v,w),t})} & \text{ if } \Delta_C > 0,
    \end{cases}
\end{equation}
where $\Delta_C :=  R^\mathrm{in}_{v, t} - \sum_{(v,w) \in O_{v,t}}\min(x_{(v,w),t}, C_{(v,w),t})$.

The delay of a path $X$ with terminal node $z$ is defined as the sum of the {propagation} delays $D^\mathrm{Tx}_{(v,w),t}$ of the links $(v,w)$ in the path and the queuing delays $D^\mathrm{q}_{v,t}$ of all intermediate nodes $v$:
\begin{equation}
    D_X := \min\left(\sum\nolimits_{(v,w) \in X}{D^\mathrm{Tx}_{(v, w),t} + D^\mathrm{q}_{v, t}}, T_\mathrm{max}\right).
\end{equation}
If the terminal node in path $X$ is not the internet node $z$ or if the path contains a loop, the traffic is dropped and $D_X$ is set to $T_\mathrm{max}$. The same applies if the paths' delay exceeds $T_\mathrm{max}$.

The cost $c_{v,t}$ at satellite $v$ considers all paths in $\mathcal{X}_{v,t}$ and is calculated as the weighted sum of the delays of each path:
\begin{equation}
    c_{v,t} = \frac{\sum_{X \in \mathcal{X}_{v, t}} R_X D_X}{\sum_{X \in \mathcal{X}_{v, t}} R_X},
\end{equation}
where $R_X$ is the amount of traffic transmitted over path $X$.

The average cost $c_t$ of all satellites at $t$ is given by
\begin{equation}
\label{eq:cost}
    c_t(\mathbf{x}_t) = \frac{\sum_{n \in \mathcal{N}}R^\mathrm{g}_{n,t} c_{n,t}}{\sum_{n\in\mathcal{N}}{R^\mathrm{g}_{n,t}}} .
\end{equation}
As dropped data contributes the highest possible delay of $T_\mathrm{max}$ to \wds{$c_t(\mathbf{x}_t)$}, considering $c_t(\mathbf{x}_t)$ as the optimization target leads to a joint minimization of average delay and drop rate.
From the network perspective, $c_t(\mathbf{x}_t)$ can be minimized by solving:
\begin{equation}
    \mathbf{x}^*_t = \argmin_{\mathbf{x}_t \in (\mathbb{R}_{+})^{|\wds{\mathcal{E}_t}|}}\wds{c_t(\mathbf{x}_t)}, \quad\forall t=0,...,T-1
    \label{eq:OptProblem}
\end{equation}

Solving \eqref{eq:OptProblem} requires that every satellite has perfect knowledge of $C_{(v,w),t}$ and $R^\mathrm{g}_{v,t}$ for all other satellites $v \in \mathcal{N}\cup\mathcal{M}$, an assumption that is hard to fulfill in real systems.

%% file: sections/5_solution.tex
\section{\sysname{}{}}
\label{sec:solution}

\wds{In this section, we present \sysname{}, our proposed solution for link management in satellite networks. Unlike existing algorithms, \sysname{} is fully distributed, scalable, resilient, and does not require perfect knowledge of all $C_{(v,w),t}$ and $R^\mathrm{g}_{v,t}$ at every satellite.} We begin with an overview of its concept, followed by a detailed description and the pseudocode.

\subsection{Concept}
\label{sec:concept}
Using \sysname{}, each satellite autonomously decides which of its established links to prioritize for data transmission in order to minimize the average delay and drop rate in the network.
%This means that \sysname{} is fully distributed, with the algorithm running independently on each satellite.
\sysname{} is based on the \gls{mab} framework and uses the \gls{ucb} criterion. 
The satellite uses the \gls{mab} framework to decide which links to use. It evaluates each option using the \gls{ucb} criterion and selects its next action based on the updated evaluation.

The decision \((x_{(v,w),t} \mid w \in O_{v,t})\) of a satellite \(v\) represents the distribution of traffic across available links \(w\) in \(O_{v,t}\). Unlike traditional \gls{mab} problems with discrete choices, this decision is continuous. %, which requires the adaptation of the \gls{mab} framework.
\sysname{} addresses this by first creating a ranked list of preferences for the established links. 
Incoming traffic is then directed through the highest-ranked link on this list. 
Once its capacity is reached, additional traffic is routed through the next link in the ranking, and so forth.
\ao{\sysname{} additionally adapts its link distribution based on the particular conditions, i.e., the context, of each satellite. 
We consider that each satellite uses the distances to its neighbors as context. As a result, instead of a single global context, the satellite observes a separate context (i.e., distance) for each link. Such model allows the satellite to evaluate each link separately and based on its specific characteristics.}
%The classical \gls{mab} framework assumes the same action is optimal independent of the environment, limiting adaptability. Contextual \gls{mab} overcomes this by allowing the agent to adapt its decisions based on the context.
%For \sysname{}, each satellite uses the distances to its neighbors as context. This choice requires further adaptation of the \gls{mab} framework. Instead of a single global context, the satellite observes a separate context (i.e., distance) for each link. This allows the satellite to evaluate each link based on its specific current context.
\begin{figure}[t]
    \centering    \includegraphics[trim= 50 20 50 20,clip,width=\columnwidth]{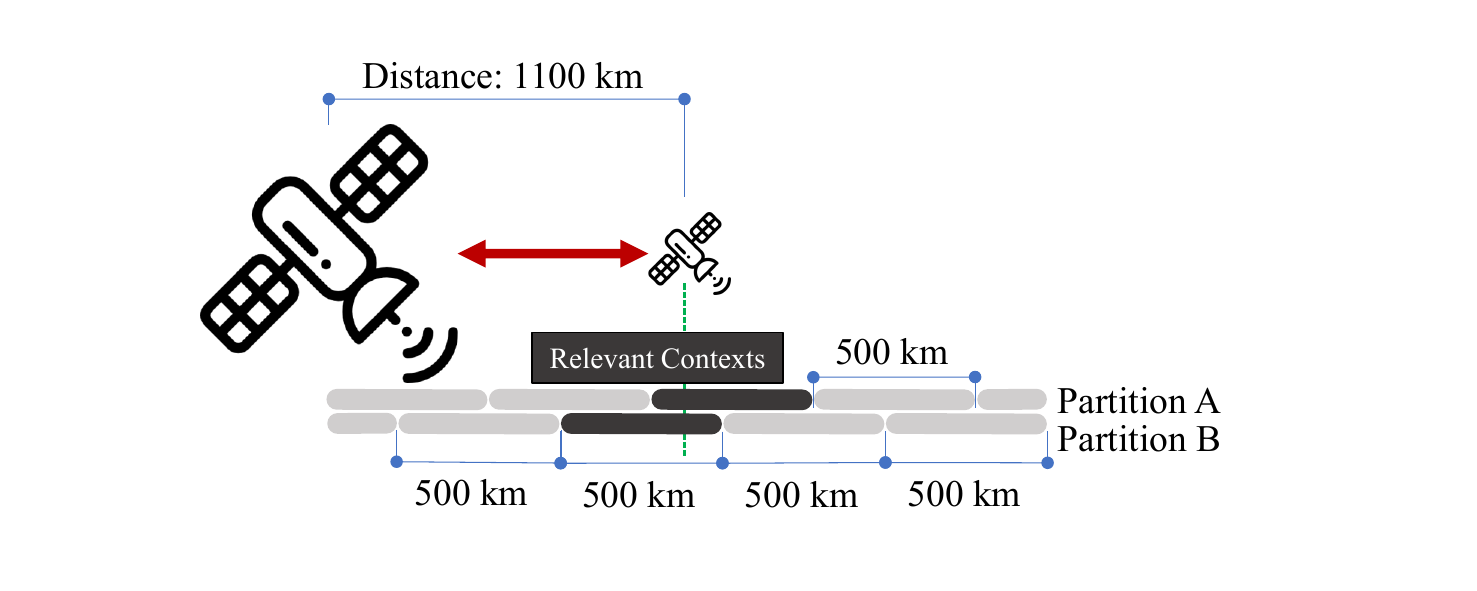}
\caption{\wds{Visualization of \sysname{}'s tile-coding mechanism.}}
    \label{fig:skylink}
\end{figure}
\ao{Note however, that the considered context is continuous. Therefore, to maintain a low computational complexity and low learning time, we quantize it into discrete partitions using a tile coding mechanism \cite{Sutton2018}. Specifically,  we divide the continuous contexts into multiple overlapping partitions. 
Under the assumption that the cost distribution is stationary in each context, our link selection algorithm inherits the convergence properties of the contextual \gls{ucb} algorithm.}
%However, because the context is continuous, it needs to be quantized into discrete categories. Contextual \gls{mab} then learn separately for each of these discrete contexts, which can result in longer learning times and abrupt shifts in decisions at the boundaries between contexts. To mitigate this problem, \sysname{} uses an additional tile coding mechanism \cite{Sutton2018}, where the continuous contexts are partitioned into multiple overlapping partitions.

\wds{In Fig. \ref{fig:skylink}, we show an example with two partitions, \wds{A and B}, which represent quantized distance ranges to \wds{a} neighbor. The satellite discretizes the distances to its neighbors into fixed-length intervals \wds{(e.g., 500 km segments; see Fig. \ref{fig:skylink})}, with a maximum distance defining the size of each partition. For every neighbor and for each context within these partitions, the satellite learns independently. When evaluating a link during decision-making, the satellite calculates the distance to its neighbors (1100 km in this example), identifies the corresponding tile (highlighted in a darker shade in the figure) within the quantized partition, and aggregates the knowledge associated with these tiles.}
In the following subsection, we explain central concepts of \sysname{} in more detail.

\subsection{Ranking Links}
\ao{Each satellite uses its experience from previous time slots to rank the established links. For every link $(v,w)$, each satellite stores and updates the average cost $\bar{c}_{v, w}(g, d_{(v,w)})$ experienced when using this connection in each given context and in each partition. The average cost  $\bar{c}_{v, w}(g, d_{(v,w)})$ is updated using a running mean. Additionally, each satellite keeps track of the number $n(v, w, g, d_{(v,w)})$ of times the link $(v,w)$ is used in each context. With this information, the satellites are able to evaluate the \gls{ucb}-criterion for every partition $g$ as}
\begin{equation}
    \text{UCB}^g_t(v, w) = \wds{\bar{c}_{v, w}}(g, d_{(v,w)}) - \sqrt{\frac{2 \log(t)}{n(v, w, g, d_{(v,w)})}}.
\end{equation}
%\wds{where $\bar{c}_{v, w}(g, d_{(v,w)})$ is the average cost for link $(v, w)$ in the tile given by partition $g$ and distance $d_{(v,w)}$ and $n(v, w, g, d_{(v,w)})$ is the number of times link $(v, w)$ has been used in this tile.}

The \gls{ucb} score is then calculated as the average over all scores for each partition:
\begin{equation}
\label{eq:ucb}
   \text{UCB}_t(v, w) = \frac{1}{|\mathcal{G}|} \sum_{g \in \mathcal{G}}\text{UCB}^g_t(v, w),
\end{equation}
where $\mathcal{G}$ is the set of partitions.
Each satellite calculates this \gls{ucb} score for all its established links. The links are then ranked based on these scores, with lower scores indicating a higher preference. The ranking determines the order in which the satellite routes and distributes incoming traffic. By continuously updating the delays and counts for each context, the satellites dynamically adjust their preferences based on the observed delays.

\subsection{Distributing Traffic}
Given the list of preferences, \sysname{} distributes the incoming traffic to its outgoing links following a water-filling pattern.
\ao{Each satellite first estimates the capacity $C_{(v,w),t}$ of each \gls{isl} and \gls{gsl} link $(v,w)$, as described in Sec. \ref{sec:systModel}}. %using Eq. \ref{eq:capacity_isl} or Eq. \ref{eq:capacity_gsl}, respectively. 
To account for uncertainties concerning the capacity, for example,  the uncertainty caused by the fact that the atmospheric attenuation $A_{\text{atmos}}$ is weather-dependent, we include an uncertainty factor $\sigma <1$ when determining the link's capacity. By using $\sigma C_{(v,w),t}$, \sysname{} ensures that the capacity of a link is not overestimated, which would result in higher drop rates.

% \ao{After estimating the link capacities, the decision of satellite $v$ is given as follows. Assume that the incoming data rate at satellite $v$ at time slot $t$ is given by $R^{\text{in}}_{v, t}$ and the list of preferences for its established links is given by 
\ao{After estimating the link capacities, the decision of satellite $v$ is given as follows. For an incoming data rate $R^{\text{in}}_{v, t}$ at time slot $t$, let the list of preferences for its established links be given by $(w_1,...,w_k)$, $k=|O_{v,t}|$. 
%Moreover, let the corresponding estimated capacities be denoted as $(C_1,...,C_k)$. 
%Further assume that for some $i < k$}
Now assume that for some $i < k$}
\ao{\begin{equation*}
    \sum_{j=1}^i \sigma C_{(v,w_j),t} < R^{\text{in}}_{v, t},
    \end{equation*}
    and
    \begin{equation*}
    \sum_{j=1}^{i+1} \sigma C_{(v,w_j),t} \geq R^{\text{in}}_{v, t}.
\end{equation*}
Then the full capacity of the first $i$ links is used, i.e.,
% \ao{\begin{equation*}
%     \sum_{j=1}^i \sigma C_j < R^{\text{in}}_{v, t} \quad \text{ and } \quad \sum_{j=1}^{i+1} \sigma C_j \geq R^{\text{in}}_{v, t}.
% \end{equation*}}
\begin{equation}
\label{eq:waterfilling1}
    x_{(v,w_j),t} = \sigma C_j, \quad j=1,...,i.
\end{equation}}
For link $i+1$, \sysname{} sets:
\begin{equation}
\label{eq:waterfilling2}
    x_{(v,w_{i+1}),t} = R^{\text{in}}_{v, t} - \sum_{j=1}^{i} \sigma C_j.
\end{equation}
\revision{Within any used outgoing link, capacity is divided proportionally among all incoming streams. Through this flow-level proportional sharing, we approximate the behavior of packet schedulers.}
The remaining links $j=i+2,...,k$ are not used and $x_{(v,w_{j}),t} = 0$.
If the incoming traffic exceeds the sum of the capacity of all established links, each link is used to its full capacity and the remaining data in the buffer is dropped.

\subsection{Implementation}
\input{sections/5a_algorithm}

We describe how \sysname{} is implemented by explaining the pseudo-code shown in Alg. \ref{alg:algo}.
Note that the pseudo-code is provided for a single satellite $v$ at time slot $t$. Each of the satellites simultaneously executes this algorithm.

In every time slot $t$, \sysname{} observes the satellite's neighbors $(w_i)^k_{i=1}$ to which $v$ is connected either by an \glspl{isl} or by a \glspl{gsl} and calculates the \gls{ucb}-scores for each link using tile coding as described in Eq. \ref{eq:ucb} (lines 1-6). Based on these scores, \sysname{} creates the list of preferences (line 7).
Using this list, $v$ decides on the amount of traffic it relays through each of its links according to the water-filling mechanism described in Eq. \ref{eq:waterfilling1} and Eq. \ref{eq:waterfilling2} (lines 8-15).

Finally, \sysname{} observes the delay resulting from its decision (line 16) and updates the counter and the average delay of its preferred link (lines 17-21).

\revision{We based our implementation of \sysname on distances as contexts, because experiments showed that, among all tested features, per-link distance has the most direct and strongest influence on routing performance, outperforming any single alternative and all feature combinations. Beyond per-link distance, we evaluated local and UTC time, satellite load, neighbor distance, total path distance, and their combinations as contexts.}

\subsection{Computational Complexity}
\label{sec:complexity}
\revision{We analyze the time complexity of \sysname per satellite and per time slot. Here, $k = |O_{v,t}|$ denotes the number of established outgoing neighbors (\glspl{isl} and \glspl{gsl}), and $|\mathcal{G}|$ is the number of tile-coding grids.
For each neighbor and each grid, \sysname{} performs a constant-time update.
Each satellite establishes at most four \glspl{isl} to neighboring satellites. Additional neighbors are the visible ground stations with which \glspl{gsl} are established. Hence $k \le 4 + \big|\text{visible \glspl{gsl} at }t\big|$ and is independent of constellation size.
Ranking the $k$ neighbors is in  $\mathcal{O}\!\big(k\log k\big)$.
The water-filling loop performs at most $\sum_{i=1}^k i = \mathcal{O}(k^2)$ operations. With prefix sums, this reduces to $\mathcal{O}(k)$.
Finally, updating the running means is linear in the number of grids. 
Overall, per satellite and time slot, the time complexity of \sysname{} is in $\mathcal{O}\!\big(k\,|\mathcal{G}| + k\log k\big)$.
This makes the complexity independent of the number of satellites and users because $k$ is bounded by visibility of ground stations rather than constellation size or demand.
}

%% file: sections/5a_algorithm.tex
\begin{comment}
\begin{algorithm}[t]
\caption{\sysname{} at satellite $v$}\label{alg:algo}
\begin{algorithmic}[1]
\doublespacing
\footnotesize
% Simulation
\For{$t \in \{0,..., T-1\}$}
    \State{Observe connected neighbors $(w_i)^k_{i=1}$ of $v$}
    \State{Sort $(w_i)^k_{i=1}$ according to $\text{UCB}_t(v,w_i)$: \Comment{Eq. \eqref{eq:ucb}}
    $$ \text{UCB}^g_t(v, w) = \bar{D}_{v, w}(g, d_{(v,w)}) - \sqrt{\frac{2 \log(\hat{N}_v)}{n(v, w)}} $$}
    \State{Set $(x_{(v,w_i),t})^k_{i=1}$: \Comment{Eq. \eqref{eq:waterfilling1}, Eq. \eqref{eq:waterfilling2}}:
    $$ x_{(v,w_j),t} = C_j, $$
    $$ x_{(v,w_{i+1}),t} = R^{\text{in}}_{v, t} - \sum_{j=1}^{i} C_j. $$}
    \State Observe delay $D_{v, t}$ 
    \State Update $\bar{D}_{v, w_1}(g, d_{(v,w_1)})$, $n(v,w_1)$ for every $g \in \mathcal{G}$.
\EndFor
\end{algorithmic}
\end{algorithm}
\end{comment}

\begin{algorithm}[t]
\caption{\wds{\sysname{} at satellite $v$ in time slot $t$}}
\label{alg:algo}
\begin{algorithmic}[1]
\footnotesize
\doublespacing
    
\For{$w_i \in O_{v,t}$}: \Comment{Calculate UCB-value, Eq.~\eqref{eq:ucb}}
    \For{$g \in \mathcal{G}$}:
        \State $\text{UCB}^g_t(v, w_i) = \bar{c}_{v, w_i}(g, d_{(v,w_i)}) - \sqrt{\frac{2 \log(t)}{n(v, w_i, g, d_{(v,w_i)})}}$
    \EndFor
    \State $\text{UCB}_t(v, w_i) = \frac{1}{|\mathcal{G}|} \sum_{g \in \mathcal{G}}\text{UCB}^g_t(v, w_i)$
\EndFor

\State \textbf{Sort} neighbors $(w_i)_{i=1}^k$ in ascending order of $\text{UCB}_t(v, w_i)$

\For{$i = 1$ \textbf{to} $k$} \Comment{Allocate flow, Eqs.~\eqref{eq:waterfilling1},~\eqref{eq:waterfilling2}}
    \If{$\sum_{j=1}^{i} C_j < R^{\text{in}}_{v,t}$}
        \State $x_{(v, w_i), t} \gets C_i$
    \Else
        \State $x_{(v, w_i), t} \gets R^{\text{in}}_{v,t} - \sum_{j=1}^{i-1} C_j$
        \State \textbf{break}
    \EndIf
\EndFor

\State \textbf{Observe} cost $c_{v,t}$
\For{$g \in \mathcal{G}$}:
    \State $n \gets n(v, w_1, g, d_{(v,w_1)})$
    \State $\bar{c}_{v, w_1}(g, d_{(v,w_1)}) \gets \frac{n \cdot \bar{c}_{v, w_1}(g, d_{(v,w_1)}) + c_{v,t}}{n + 1}$
\EndFor
\State $n(v, w_1, g, d_{(v,w_1)}) \gets n(v, w_1, g, d_{(v,w_1)}) + 1$

\end{algorithmic}
\end{algorithm}

%% file: sections/6_simulator.tex
\section{LEO Network Simulator}
\label{sec:simulator}

Existing simulation tools for the evaluation of link management in \gls{leo} satellite networks are usually limited to network segments (see e.g. \cite{Roth2024, Deng2023}). Therefore, to validate our proposed approach at a global scale and support the advancement of \gls{leo} satellite networks, we present a novel simulation framework.
Within our simulator, we make use of CosmicBeats \cite{CosmicBeatsSimulator} to pre-calculate the exact position of each ground stations and each satellite in every time slot over the time horizon.
To perform these calculations, CosmicBeats uses a list of coordinates for each ground station, \gls{tle}-data for each satellite in the considered constellation, a time slot duration $\tau$ and a time frame including a starting time and a number of time slots $T$.
\revision{We use a flow-level (stream-based) simulator. As described in Sec. \ref{sec:systModel}, user data are represented as continuous rates per time slot. Individual packets are not instantiated.}
Based on this information, our simulator enables global-scale evaluation of link management approaches by combining the following components, which we describe in detail in the following.

\textbf{Grid Construction}
For each satellite, we pre-calculate its direct neighboring satellites in the constellation grid and the visible ground stations it can connect to. Using the satellite's location information from CosmicBeats, the simulator builds a grid containing at most four close, visible neighbor satellites (one in each cardinal direction), as well as the visible ground stations in each time slot. \ao{The grid is limited to four neighbors to account for the number of available \gls{isl} links.}

\textbf{Channel Models}
\ao{Using the satellites' positions and ground stations' locations, the simulator determines, in each time slot, the channel conditions based on the models described in Sec. \ref{sec:systModel}.  With this information, the capacities $C_{(v,w),t}$ of the \glspl{isl} and \glspl{gsl} are calculated. Note that we pre-calculate the atmospheric attenuation $A_{\text{atmos}}$ for different angles between satellites and ground stations}.

\textbf{Data Generator}
\wds{The simulator uses a global population dataset \cite{populationDataset} to assign each satellite an amount of traffic proportional to the population living close to its position. The amount of traffic is then calculated as detailed in Eq. (\ref{eq:userdata}).}

\textbf{Data Stream Transmission}
Using the generated network grid, its channel properties, and the incoming rates, we simulate the transmission of data streams in the network. 
As a result, we obtain the experienced delay, drop rate, and throughput per satellite, as well as network metrics like total network throughput and average number of hops per path.

\textbf{Visualization}
\begin{figure}
    \centering
    \includegraphics[width=\columnwidth]{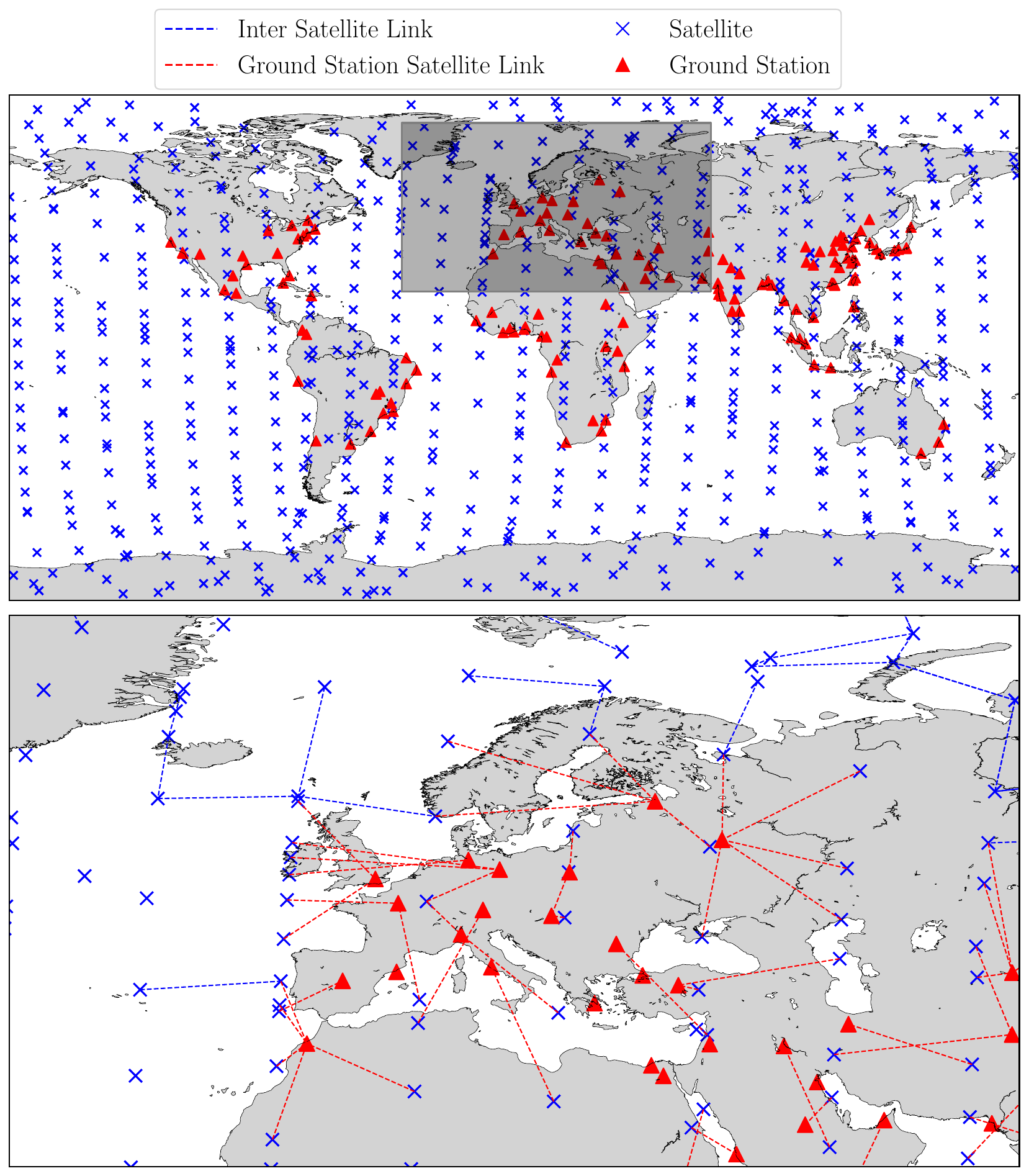}
    \caption{The simulated network including OneWeb satellites and ground stations. The section at the bottom is highlighted in the top map and includes \glspl{gsl} and \glspl{isl} carrying traffic.}
    \label{fig:europe}
\end{figure}
Our simulator offers multiple options for visualizing both the network structure and the performance of various routing strategies. An example of a simulated network visualization is shown in Fig.~\ref{fig:europe}. The upper portion depicts the entire OneWeb \gls{leo} network as of September 26, 2023, at 08:51:00 UTC. In the lower portion, a zoomed-in section of the network is presented, additionally including \glspl{isl} and \glspl{gsl} that are actively carrying traffic. %Network performance evaluation and visualization is presented in the following section.

%% file: sections/7_results.tex
\section{Evaluation}
\label{sec:evaluation}
\input{sections/7a_parameters}
In this section, we evaluate \sysname{} and compare its performance to reference schemes using data from OneWeb's \gls{leo} satellite network. We assess the system's performance across various operational scenarios and conditions. We first explain the setup and the used parameters. Second, we introduce the reference schemes. Third, we demonstrate its scalability by evaluating performance across varying numbers of users. We then show \sysname's resilience by analyzing the impact of network failures. Finally, we present a short discussion on the parameter optimization of \sysname.

\textbf{Setup and Parameters:}
To evaluate the performance of \sysname and to compare it to the performance of reference schemes, we employ our simulator described in Sec. \ref{sec:simulator}. 
Table \ref{tab:simulation_parameters} provides an overview of the simulation parameters. 
We repeated all experiments for $R=100$ times and present an average over these repetitions. We use OneWeb's \gls{leo} satellite network with a near-polar Walker Star configuration. Ground stations are placed in the world's $M=146$ largest cities, which reflects the current scale of network infrastructures \cite{GroundstationStatistics}. For the simulated time frame, $N=636$ OneWeb satellites were operational. OneWeb's \gls{tle} data is available in \cite{celestrak_oneweb_tles}.
To generate the incoming data, we use the global population dataset in \cite{populationDataset}. 
We assume 25 million users globally out of $8$ billion people (i.e., $\mathrm{d} = 3.175 \cdot 10^{-3}$), reflecting the high annual growth and recent statistics. We estimate the annual amount of generated traffic as $6$ Zettabytes \cite{ITU2023}, where upload traffic accounts for $8\%$ of the total traffic \cite{Ericsson2023TrafficProfiles}. 
Dividing by the number of seconds per year and a total of $5.3$ billion fixed and mobile broadband network users \cite{CiscoAnnualInternetReport}, we set $\nu = 22.98$ kbps.
The generated traffic per second varies depending on the local time. $\mathrm{t}_{n,t}$ is adjusted every hour, matching a characteristic daily pattern observed from \cite{AMSIX2024}.

\textbf{Reference Schemes:}
\revision{To ensure a fair comparison, we include only baselines whose signaling overhead is not higher than \sysname’s fully distributed design, which uses only locally available information. We exclude schemes that require exchanging state information or a central controller.
Concretely, we compare \sysname to five reference schemes, each highlighted with a distinct color in the upcoming figures for clarity:}
$(i)$ a bent-pipe solution that sorts only the established \glspl{gsl} randomly for its preference list (\textcolor{customorange}{orange}), 
$(ii)$ a solution based on Dijkstra's algorithm that considers only the shortest path to the ground (\textcolor{customgray}{gray}), and 
$(iii)$ a solution using not just the shortest, but the $k$-shortest paths to the ground \cite{Gounder1999}, with $k=4$ (\textcolor{customgreen}{green}).
$(iv)$ A random solution that sorts the established links randomly to generate its preference list (\textcolor{customblue}{blue}),
\revision{$(v)$ A solution based on distributed Q-learning as proposed in \cite{Soret2023} (\textcolor{custombeige}{beige}).}

In addition, we evaluate \sysname against a variant, non-contextual \sysname (NC-\sysname), which omits both contextualization and the tile-coding mechanism, and is represented in \textcolor{custompurple}{purple}.

\begin{figure*}[htb]
    \centering
    \begin{subfigure}[b]{0.32\textwidth}
        \centering
        \includegraphics[width=\linewidth]{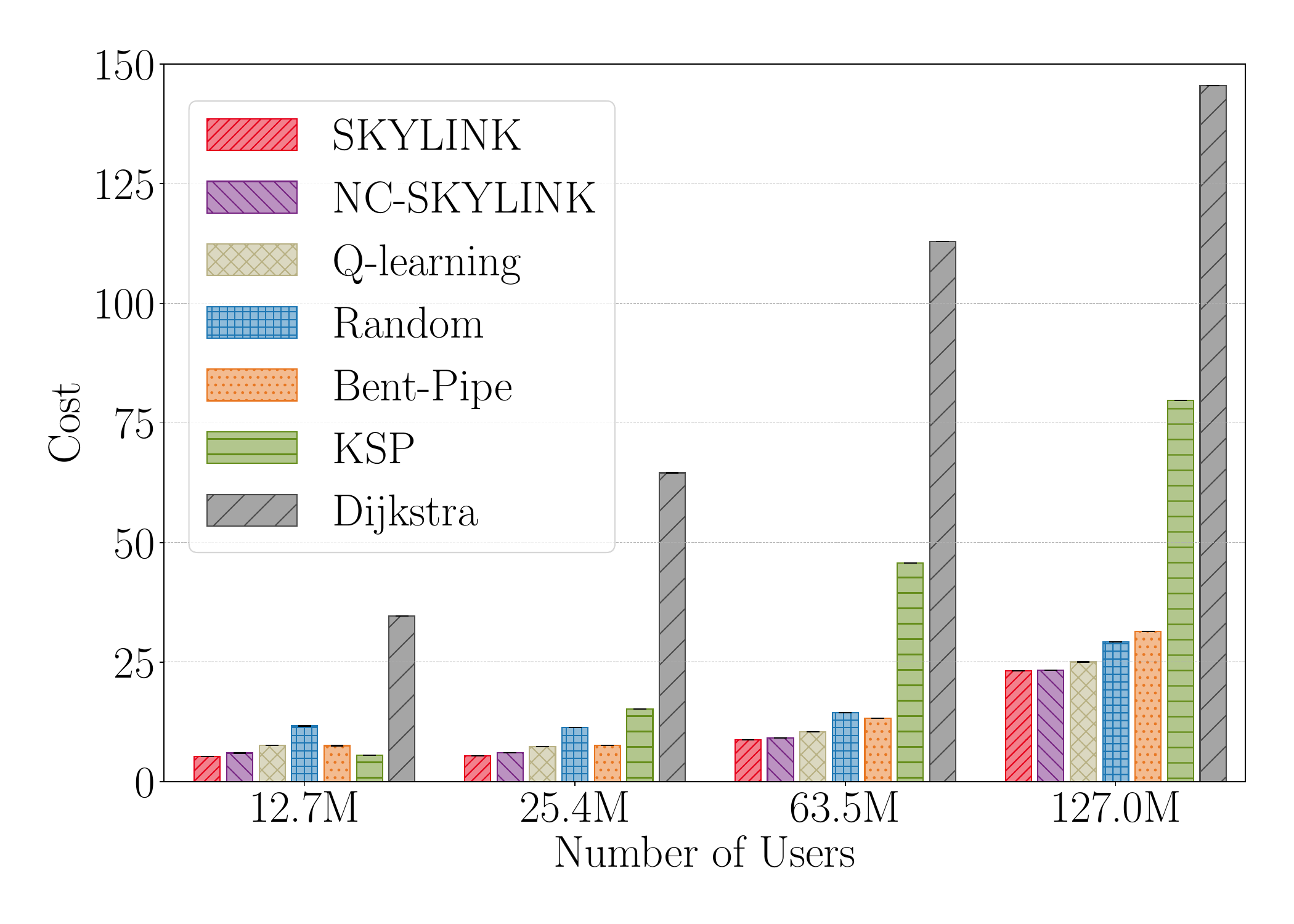}
        \caption{Average Cost.}
        \label{fig:barplot_cost}
    \end{subfigure}
    \begin{subfigure}[b]{0.32\textwidth}
        \centering
        \includegraphics[width=\linewidth]{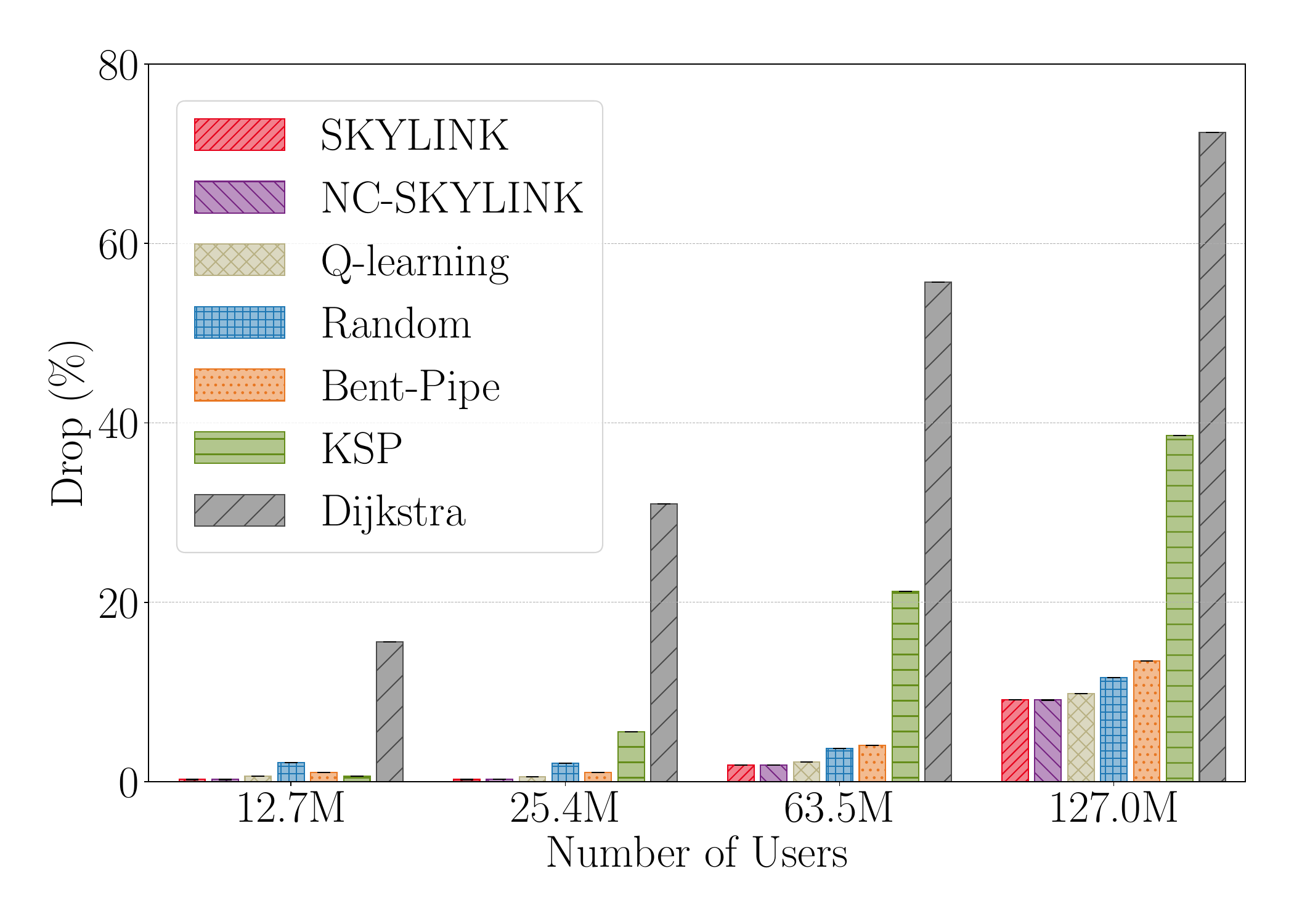}
        \caption{Average Drop Rate.}
        \label{fig:barplot_drop}
    \end{subfigure}
    \begin{subfigure}[b]{0.32\textwidth}
        \centering
        \includegraphics[width=\linewidth]{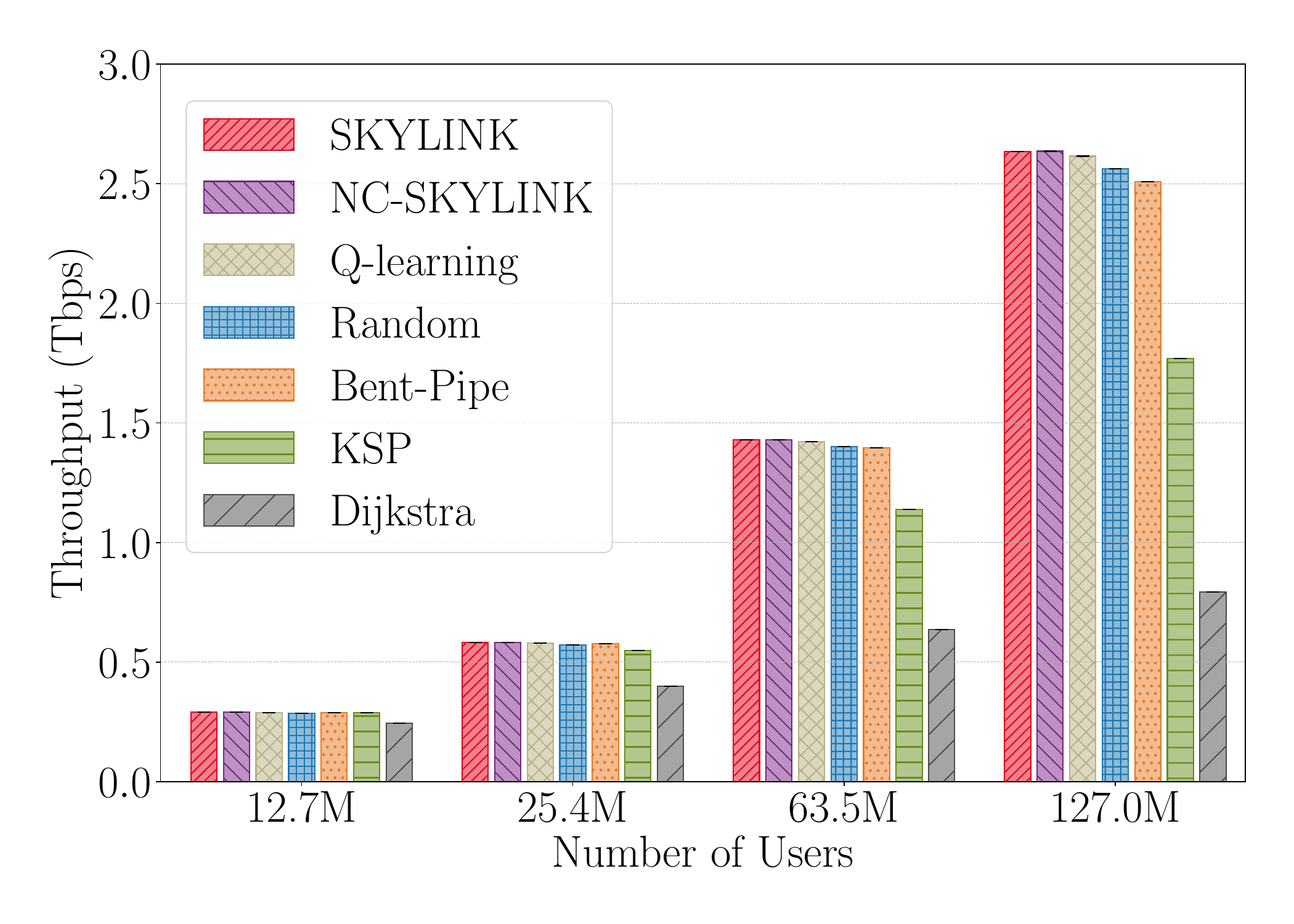}
        \caption{Total Throughput.}
        \label{fig:barplot_tp}
    \end{subfigure}
    \caption{\revision{Comparison of \sysname and the reference schemes for different metrics and different numbers of users.}}
    \label{fig:barplots}
\end{figure*}
\begin{figure*}[htb]
    \centering
    \begin{subfigure}[b]{0.49\textwidth}
        \centering
        \includegraphics[width=\columnwidth]{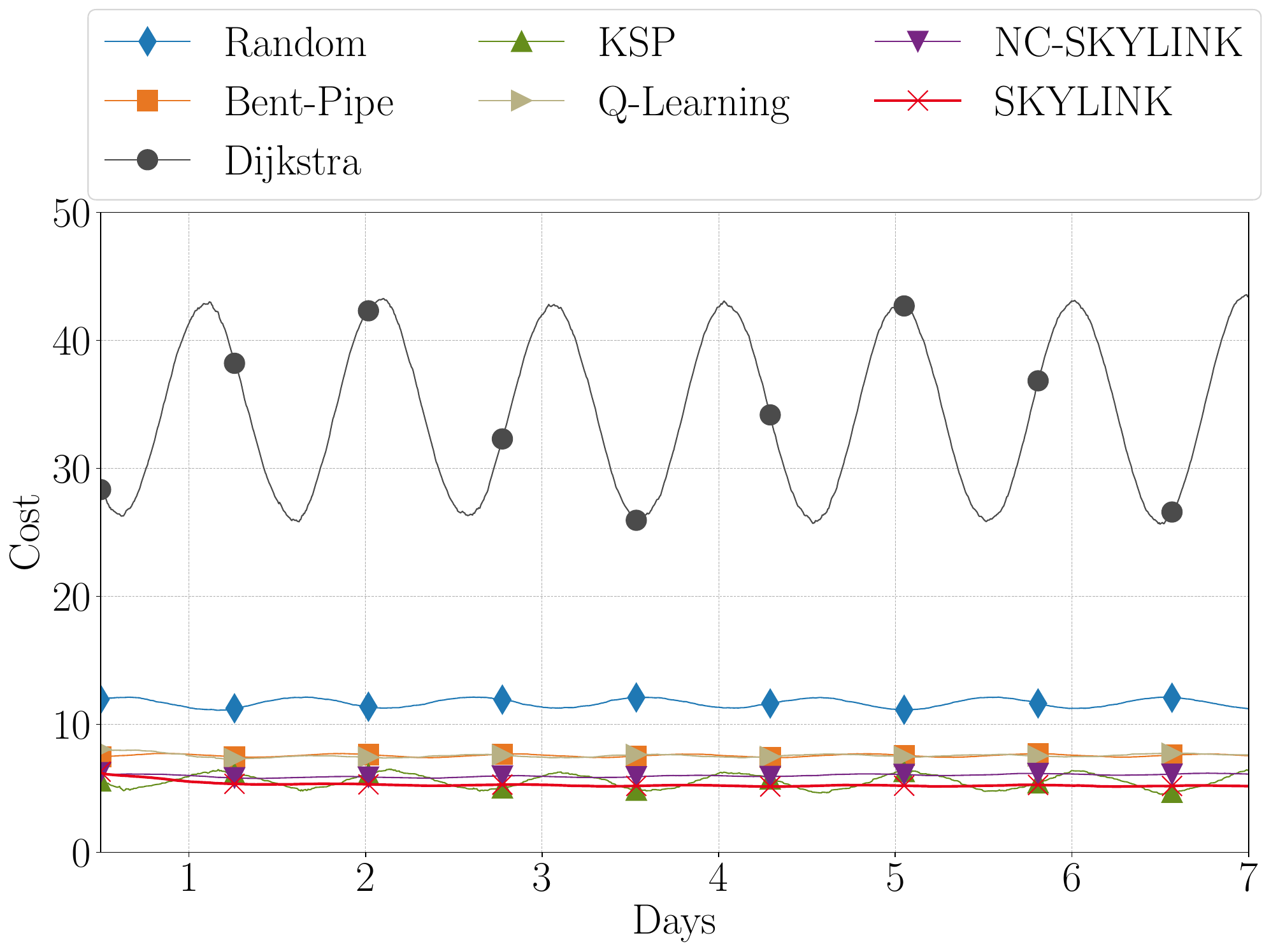}
        \caption{Cost over time for $12.7$ million users.}
        \label{fig:timeplot_cost_13M}
    \end{subfigure}
    \hfill
    \begin{subfigure}[b]{0.49\textwidth}
        \centering
        \includegraphics[width=\columnwidth]{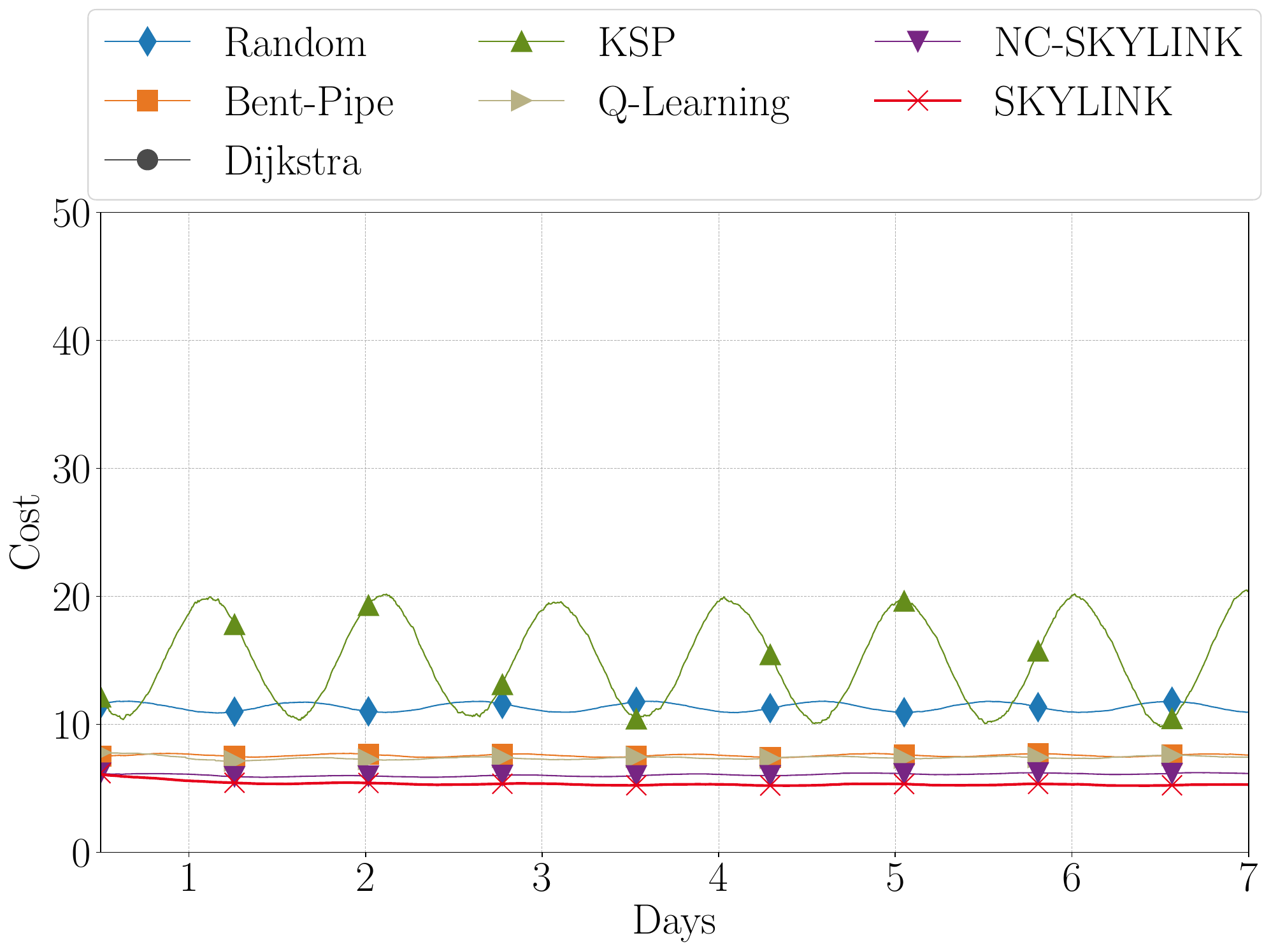}
        \caption{Cost over time for $25.4$ million users.}
        \label{fig:timeplot_cost_25M}
    \end{subfigure}
    \caption{\revision{Evolution of cost over a week for different user scales.}}
    \label{fig:timeplots_cost_scalability}
\end{figure*}

\subsection{Scalability}
\begin{figure*}[htb]
    \centering
    \begin{subfigure}[b]{0.49\textwidth}
        \centering
        \includegraphics[width=\columnwidth]{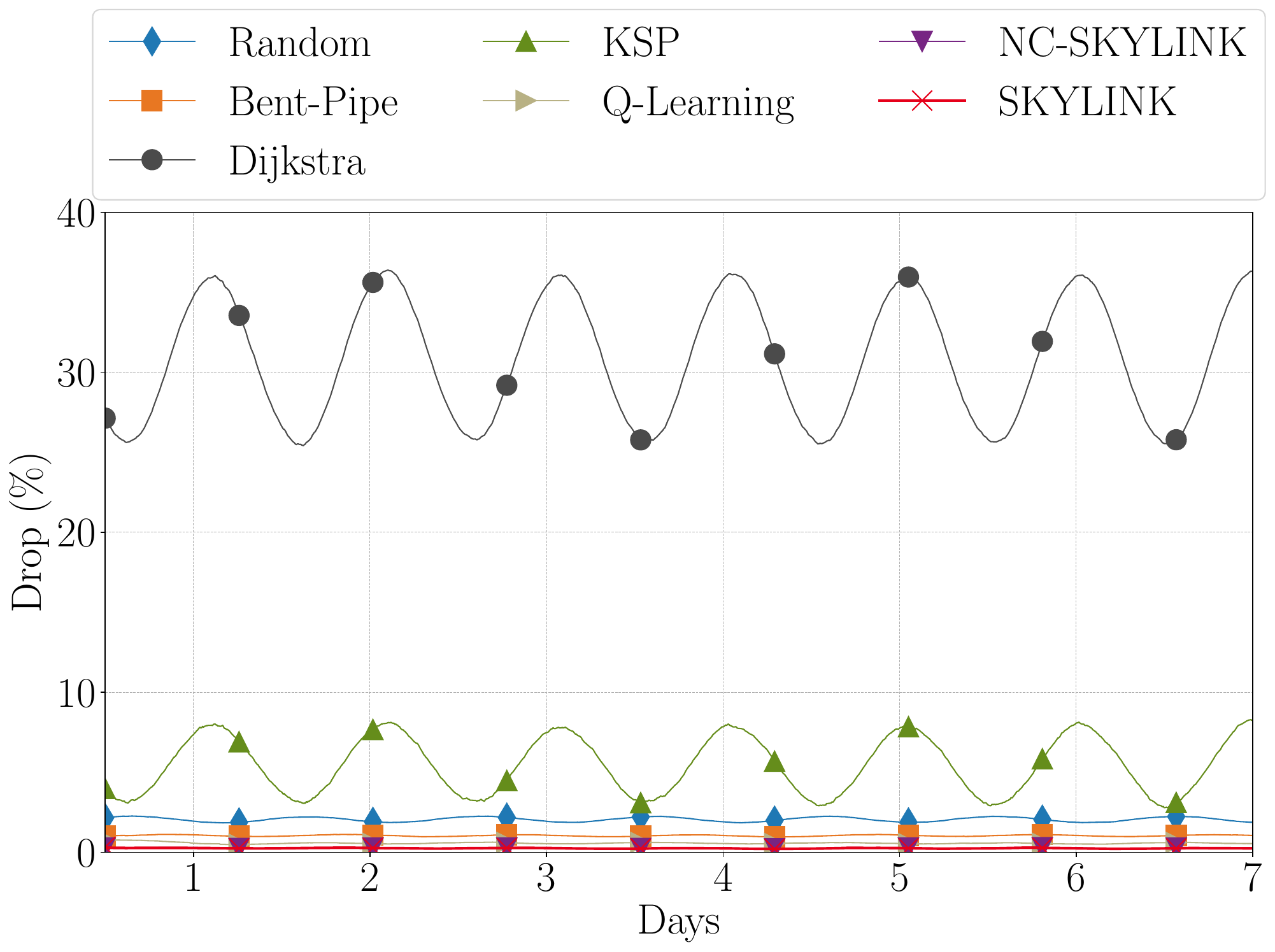}
        \caption{Drop rate over time.}
        \label{fig:timeplot_drop}
    \end{subfigure}
    \hfill
    \begin{subfigure}[b]{0.49\textwidth}
        \centering
        \includegraphics[width=\columnwidth]{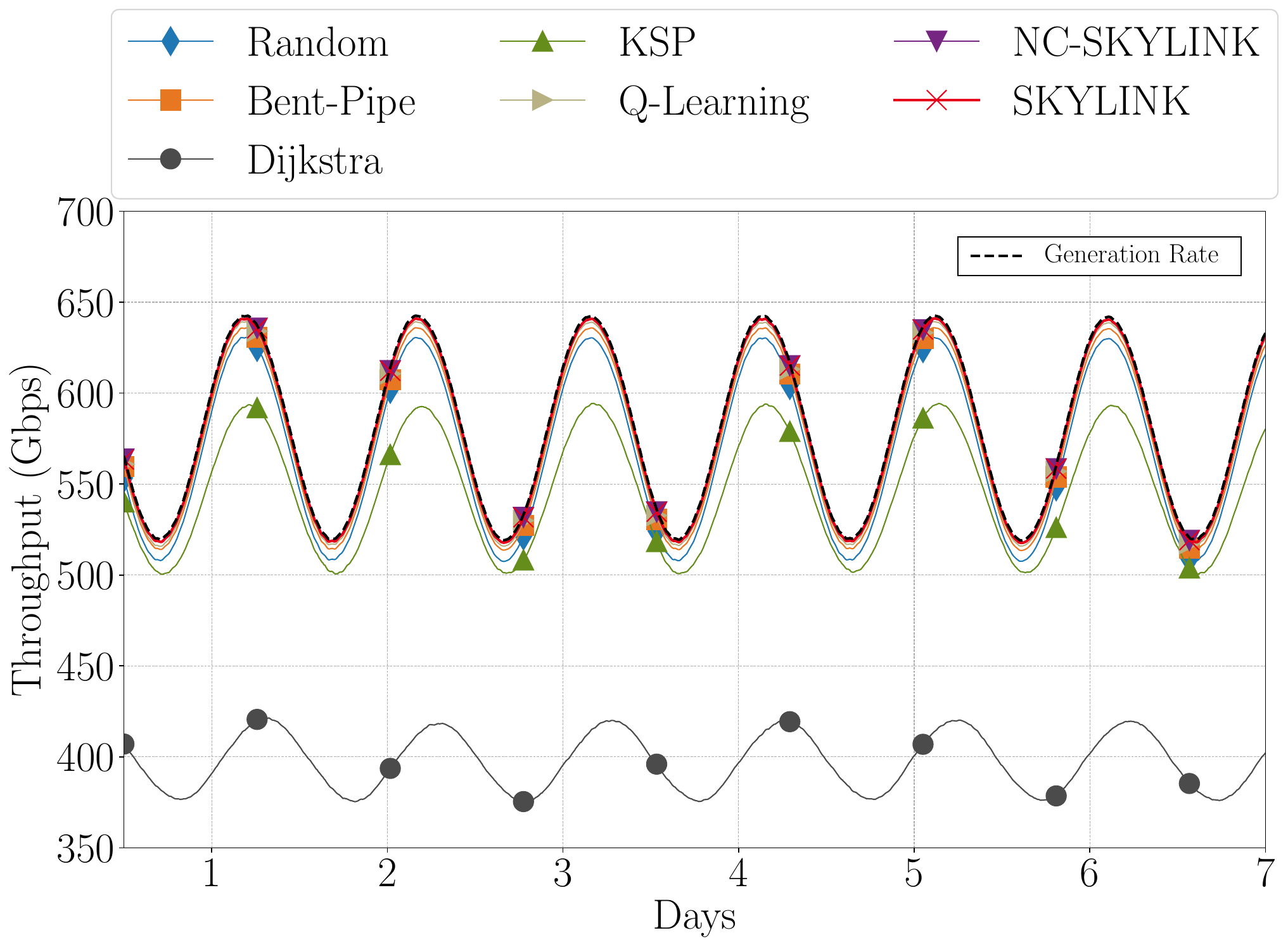}
        \caption{Throughput over time.}
        \label{fig:timeplot_tp}
    \end{subfigure}
    \caption{\revision{Evolution of cost and throughput over a week for $25.4$ million users.}}
    \label{fig:timeplots_drop_tp}
\end{figure*}
\begin{figure}
    \centering
    \includegraphics[width=\columnwidth]{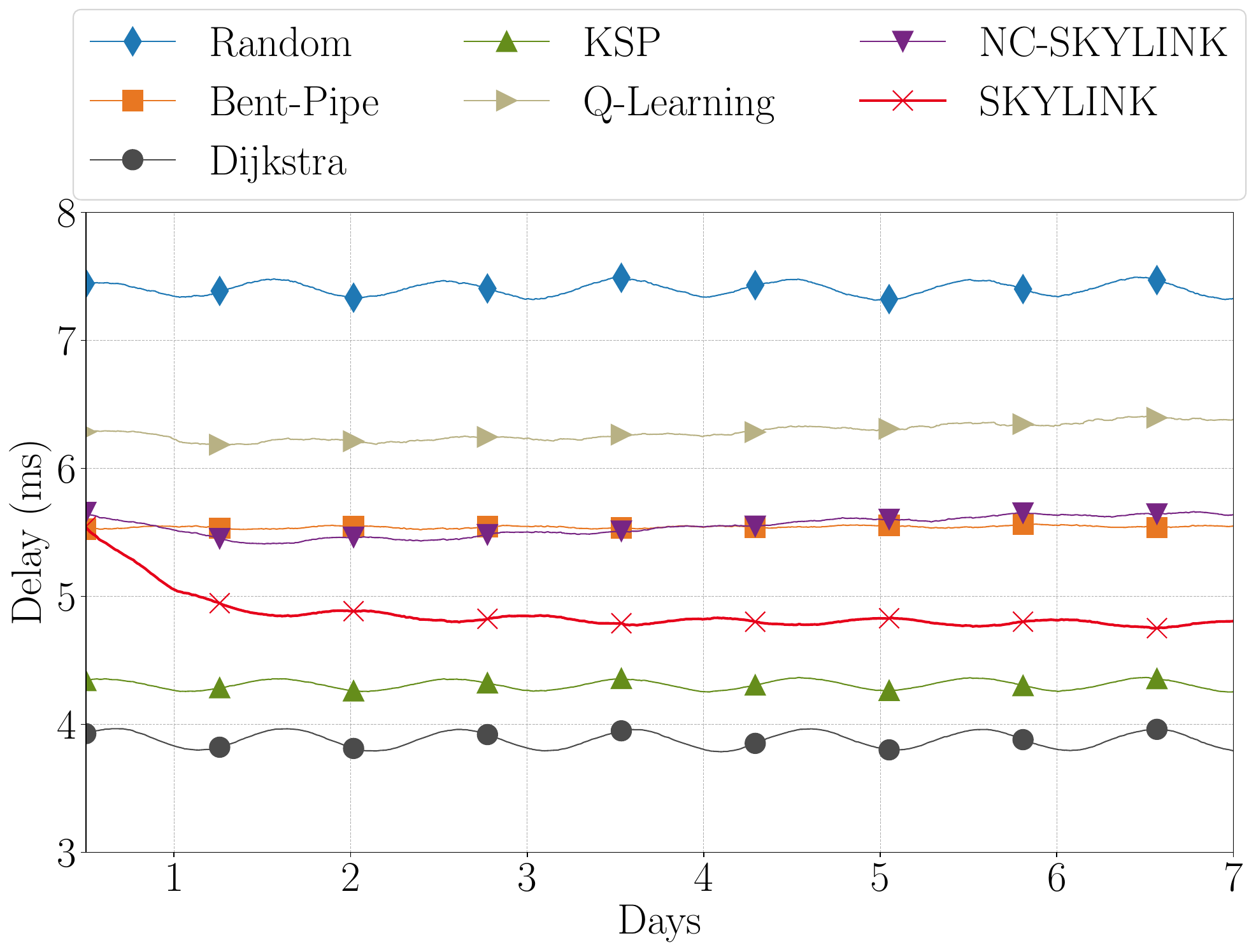}
    \caption{\revision{Evolution of delay over a week for $25.4$ million users.}}
    \label{fig:timeplot_delay}
\end{figure}

We begin by demonstrating the scalability of \sysname. In Fig. \ref{fig:barplots}, we present the results of \sysname and the reference schemes for $12.7$, $25.4$, $63.5$, and $127.0$ million users respectively for different metrics. \revision{Anchoring to publicly reported Starlink adoption, we fit a simple exponential and use the projected January 2026 levels ($12.7$ million users) as a realistic upper bound for a constellations near-term user base. We then use its 1×, 2×, 5×, and 10× multiples to test scalability.} The standard deviations across all metrics are small, consistently below $1\%$ of the average value. As a result, while error bars are included in the graphs, they remain barely noticeable. In Fig. \ref{fig:barplot_cost}, we show the average cost, which is the sum of delivered traffic weighted by its respective delay and the drop rate weighted by $T_\mathrm{max}$  as in Eq. (\ref{eq:cost}). In every scenario, \sysname generates the lowest cost and outperforms \revision{distributed Q-learning},  the random solution, the bent-pipe approach, and the algorithms based on shortest paths. The most significant improvement can be observed compared to Dijkstra's algorithm. Dijkstra's algorithm only uses the shortest path from the satellite back to the ground. This results in frequent congestion of \glspl{gsl}, higher drops and consequently in higher cost. When using not only one but $k$ shortest paths, this effect is dampened. For $12.7$ million users, the $k$-shortest paths algorithm is still performing close to \sysname. However, with an increasing number of users, it shows that $k$-shortest paths suffers from the same inefficiencies as Dijkstra's algorithm. A similar effect can be observed for the bent-pipe solution. For $12.7$ million users, \sysname reduces the cost by $5.0\%$ compared to $k$-shortest paths, $11.3\%$ compared to NC-\sysname, $29.5\%$ compared to the bent-pipe solution, \revision{$29.8\%$ compared to distributed Q-learning,} $54.4\%$ compared to the random approach, and $84.6\%$ compared to Dijkstra.

As the number of users, and consequently the data volume, increases, some \glspl{gsl} become overloaded. The random solution performs comparably well because it actively utilizes the available \glspl{isl}, leading to a more balanced distribution of traffic across multiple \glspl{gsl}. However, since the \glspl{isl} are used randomly rather than strategically, this approach reduces the drop rate but increases the delay, affecting the overall cost negatively. The improvement of \sysname compared to NC-\sysname is small but constant. The reason for this is that NC-\sysname uses the same algorithm which is also fully distributed and learns likewise. The improvement is hence solely based on the additional tile-coding mechanism. 
In terms of cost and for $25.4$ million users, \sysname improves Dijkstra by $91.7\%$, $k$-shortest path by $64.5\%$, the random solution by $52.5\%$, \revision{distributed Q-learning by $27.3\%$,} and the bent-pipe approach by $28.7\%$. Using contextualization and tile-coding improves \sysname by $11.1\%$ compared to NC-\sysname. 

We present the average drop rate and the total network throughput in Fig.~\ref{fig:barplot_drop} and Fig.~\ref{fig:barplot_tp}, respectively. Clearly, using only the shortest path results in higher drop rates that increase rapidly with the number of users. Although the $k$-shortest paths approach initially achieves a low drop rate, it increases even faster than that for Dijkstra's algorithm as the number of users grows. From $12.7$ to $127$ million users, the drop rate for Dijkstra increases from $15.6\%$ to $72.4\%$, while for $k$-shortest paths, it rises from $0.6\%$ to $38.6\%$. Consequently, the total network throughput of shortest-path algorithms falls significantly behind that of \sysname.
\revision{For $25.4$ million users, \sysname achieves a $95.4\%$ reduction in drop rate compared to $k$-shortest paths and $99.2\%$ compared to Dijkstra. It also improves over distributed Q-learning, the bent-pipe approach and the random solution by $56.1\%$, $75.6\%$ and $87.6\%$, respectively.} However, both NC-\sysname and \sysname achieve a drop rate of $0.3\%$, showing that the improvements in cost are due to the fact that \sysname achieves a lower average delay than NC-\sysname. In terms of throughput, this translates to \sysname delivering $45.8\%$ higher throughput than Dijkstra and $6.0\%$ more than $k$-shortest paths.

To analyze the performance of \sysname and the reference schemes in greater detail, we not only examine aggregate metrics over the entire simulation period but also investigate how these metrics evolve over the course of a week. Note that we display running means over half a day, which is why the horizontal axis starts at $0.5$ days. The comparison of costs over time for $12.7$ million users in Fig.~\ref{fig:timeplot_cost_13M} and $25.4$ million users in Fig.~\ref{fig:timeplot_cost_25M} provides three major insights: (a) shortest-path algorithms lack scalability, (b) \sysname has a learning phase and converges within days, and (c) \sysname is resilient to fluctuations affecting other schemes. Most strikingly, the lack of scalability in shortest-path algorithms is evident. For $12.7$ million users, both Dijkstra and $k$-shortest paths exhibit signi\-ficant cost fluctuations throughout the day, which intensify as the network becomes more congested with $25.4$ million users. Additionally, \sysname undergoes a learning phase during the initial days, fully stabilizing within a week. Finally, \sysname demonstrates resilience against the fluctuations that affect all other reference schemes, maintaining consistent performance even under varying network loads. Note that for 25.4 million users, Dijkstra consistently incurs a cost greater than 50 throughout the entire period, which is why it is not visible in Fig.~\ref{fig:timeplot_cost_25M}.

\begin{figure*}[htb]
    \centering
    \begin{subfigure}[b]{0.49\textwidth}
        \centering
        \includegraphics[width=\columnwidth]{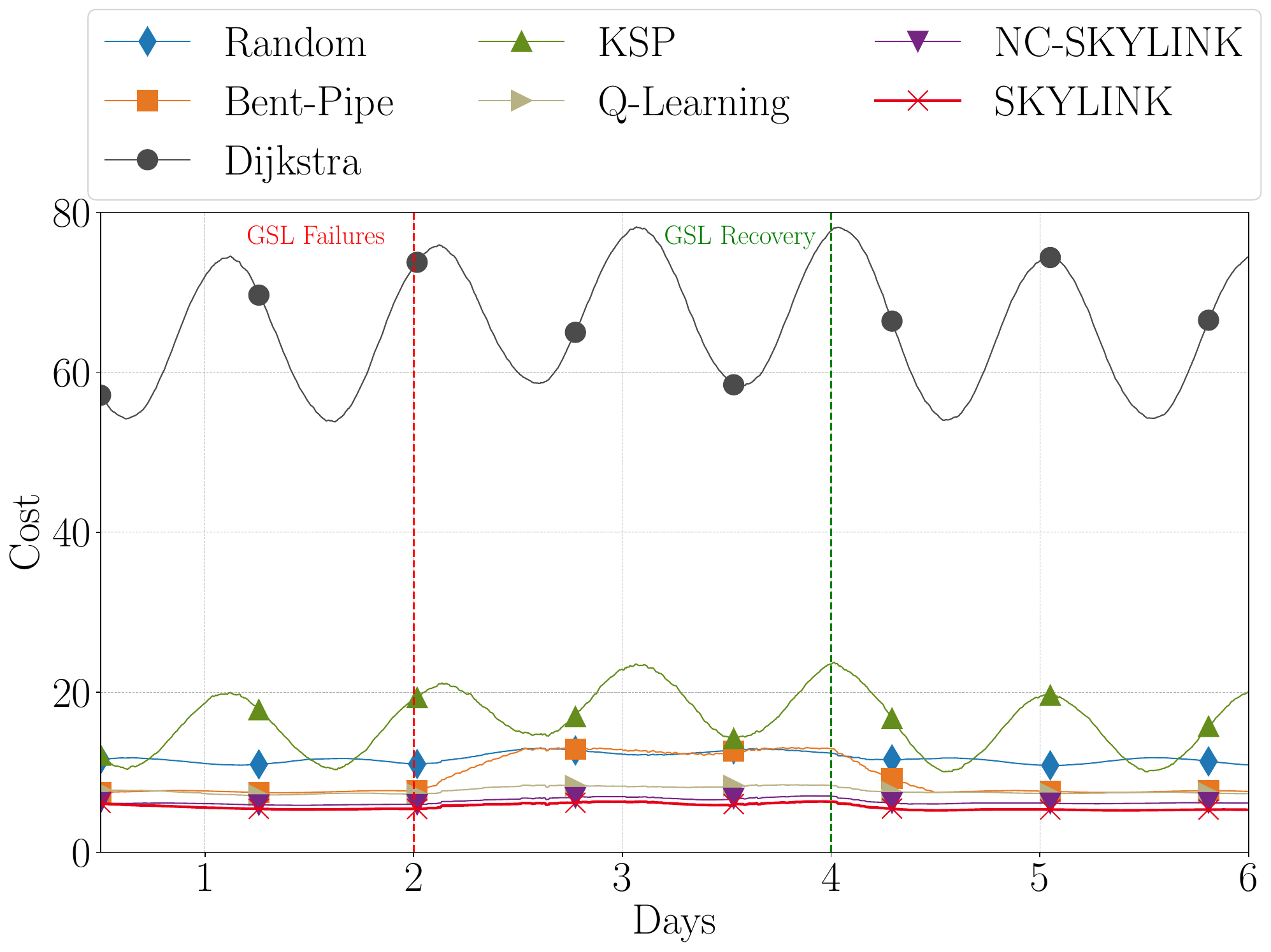}
        \caption{Cost over time.}
        \label{fig:failplot_cost}
    \end{subfigure}
    \hfill
    \begin{subfigure}[b]{0.49\textwidth}
        \centering
        \includegraphics[width=\columnwidth]{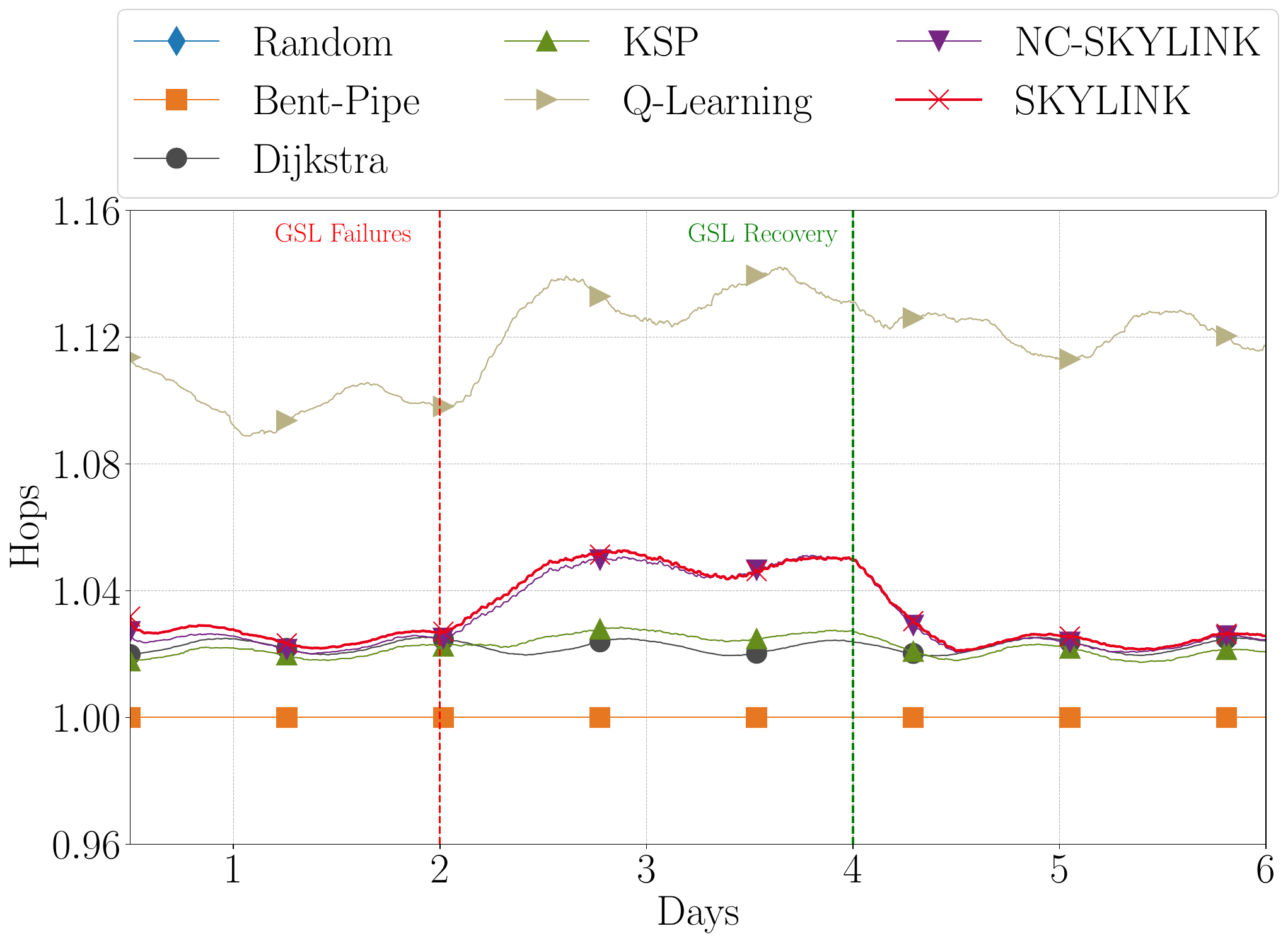}
        \caption{Average number of hops over time.}
        \label{fig:failplot_hops}
    \end{subfigure}
    \caption{\revision{Evolution of cost and average hops over a week for $25.4$ million users and under \gls{gsl}-failures.}}
    \label{fig:failplot_cost_hops}
\end{figure*}
\begin{figure*}[htb]
    \centering
    \begin{subfigure}[b]{0.49\textwidth}
        \centering
        \includegraphics[width=\columnwidth]{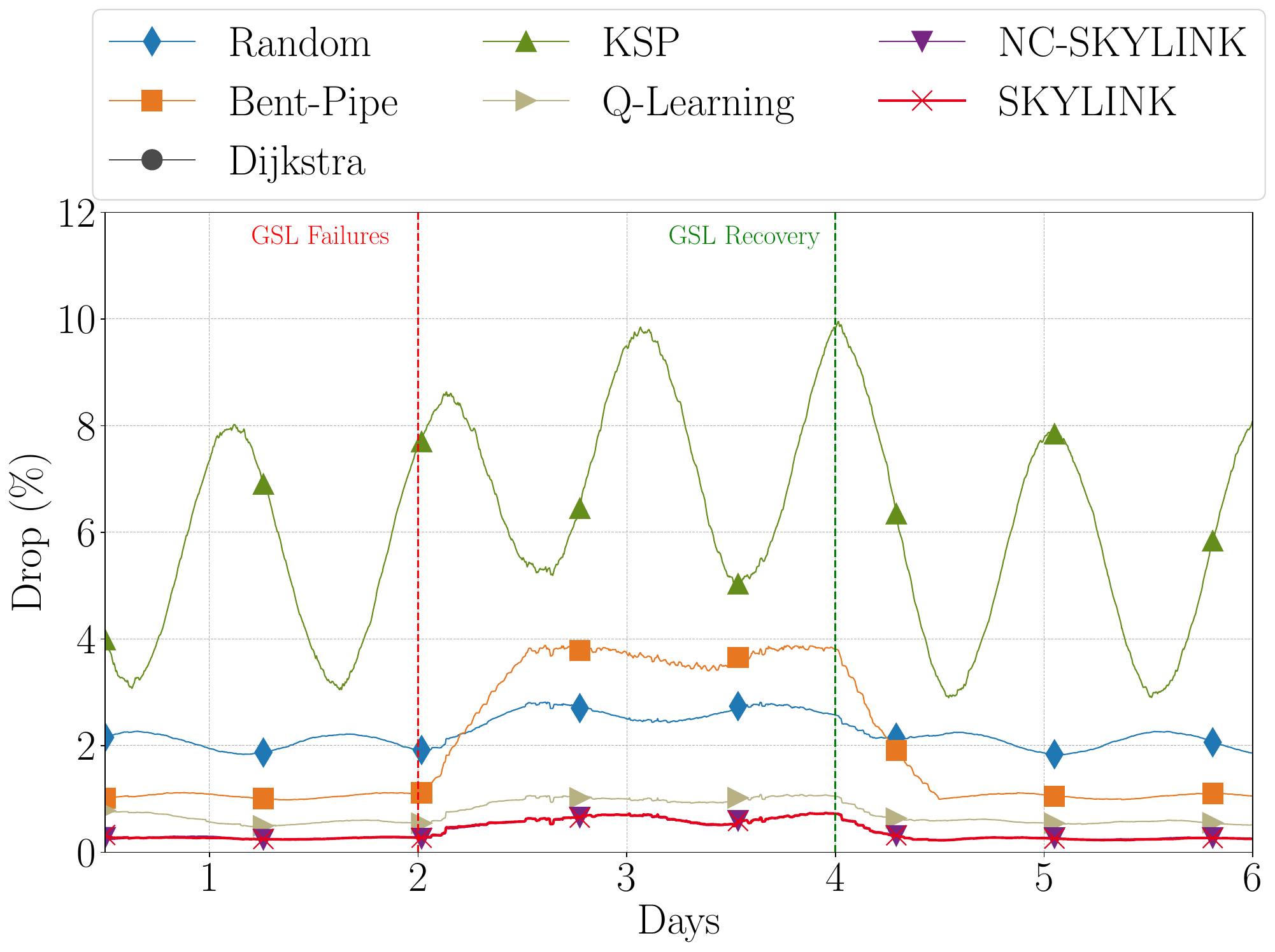}
        \caption{Drop rate over time.}
        \label{fig:failplot_drop}
    \end{subfigure}
    \hfill
    \begin{subfigure}[b]{0.49\textwidth}
        \centering
        \includegraphics[width=\columnwidth]{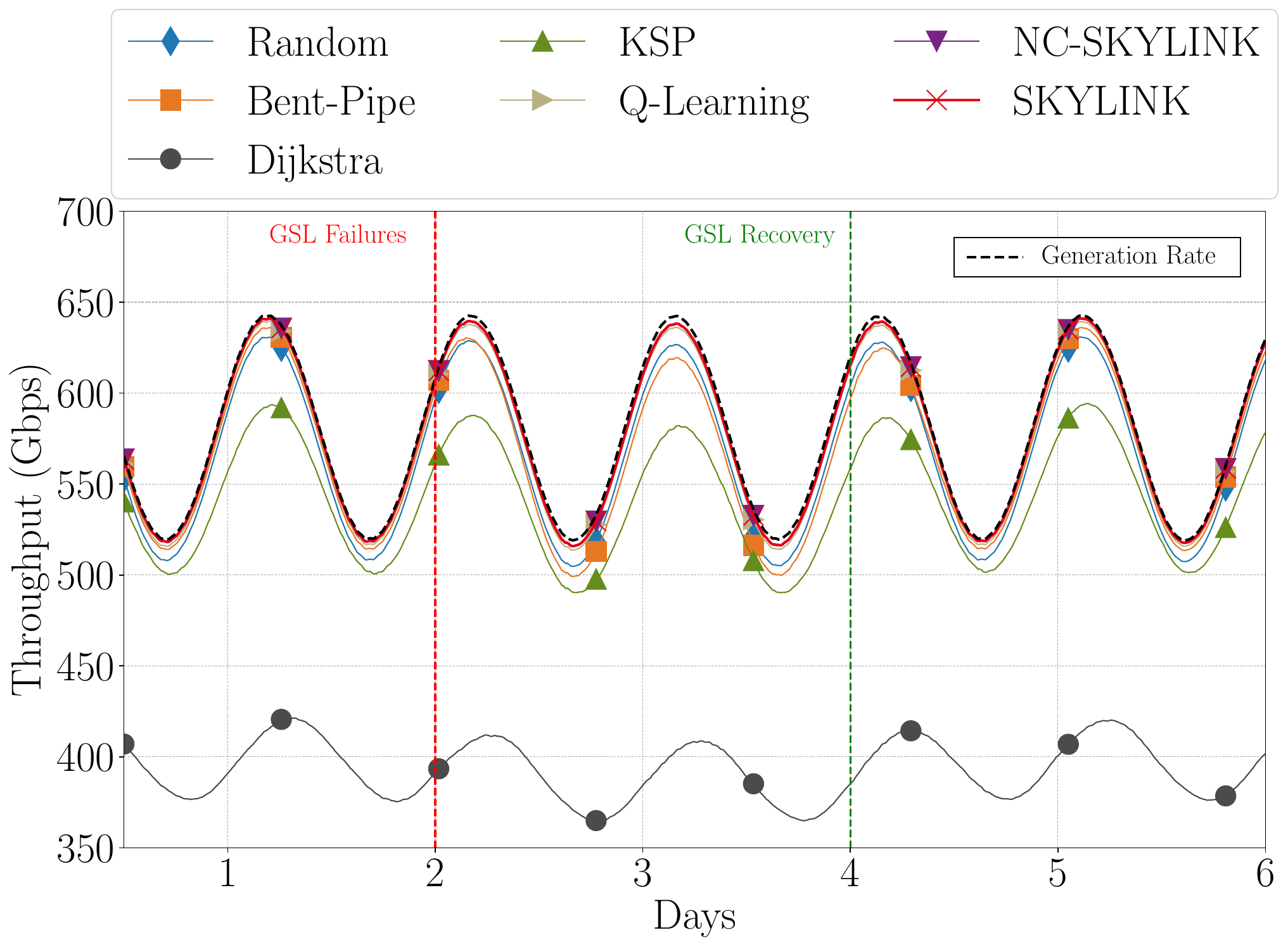}
        \caption{Throughput over time.}
        \label{fig:failplot_tp}
    \end{subfigure}
    \caption{\revision{Evolution of drop rate and throughput over a week for $25.4$ million users and under \gls{gsl}-failures.}}
    \label{fig:failplots_drop_tp}
\end{figure*}

In Fig. \ref{fig:timeplots_drop_tp}, we show the evolution of drop rate and throughput over a week for $25.4$ million users. As for the cost, in Fig. \ref{fig:timeplot_drop}, we observe a daily pattern in the drop rate resulting from the fluctuating amount of data in the network over the course of a day. The fluctuation is higher for shortest path algorithms and negligible for \sysname. In Fig. \ref{fig:timeplot_tp}, we additionally included the data generation rate in the network. As a result of the low drop rate, \sysname's total throughput is close to this generation rate. At the same time, the gap between throughput and generation rate, especially at times with high network loads is clearly visible for the reference schemes. 
For the drop rate, improvements achieved by \sysname are \revision{$56.0\%$ compared to distributed Q-learning,} $75.5\%$ compared to Bent-Pipe, $87.5\%$ compared to Random, $95.4\%$ compared to $k$-shortest paths, and $99.2\%$ compared to Dijkstra. \revision{For throughput, \sysname shows an improvement below $1.0\%$ compared to distributed Q-learning and Bent-Pipe because these strategies are able to successfully deliver almost the entire traffic.} This improvement in throughput grows to $1.8\%$ compared to Random, $6.0\%$ compared to $k$-shortest paths, and $45.9\%$ compared to Dijkstra.

As mentioned earlier, the advantage of \sysname compared to NC-\sysname is based on achieving a lower delay of delivered data. To analyze this further, we present the average delay of successfully delivered data for \sysname and the reference schemes in Fig. \ref{fig:timeplot_delay}. In particular, shortest-path algorithms, such as Dijkstra and $k$-shortest paths, exhibit relatively low delays. However, this is primarily because these algorithms deliver fewer data in general, favoring data that are closer to the ground. In contrast, \sysname delivers more data, which would be dropped by the shortest path algorithms. 
\revision{Both \sysname and NC-\sysname undergo a learning phase during the first two days. 
\sysname maintains per-link estimates conditioned on the current geometric context via distance-based tile coding, whereas NC-\sysname collapses experience across all contexts into a single global estimate. Since the relative geometry of satellites and visible ground stations continuously drifts with orbital motion, \sysname adapts its ranking to the present context, while NC-\sysname keeps averaging over mismatched situations. This leads to a gradual unlearning with a higher delay, as visible in Fig. \ref{fig:timeplot_delay}.}

\subsection{Resilience}
In the previous subsection we demonstrate that \sysname is resilient to daily fluctuations of network traffic. In this subsection, we focus on \sysname's ability to maintain its superior performance even under network failures. \revision{We restrict our analysis to \glspl{gsl} failures because they have the higher impact. Our simulations show that, even under a very unlikely scenario in which $50\%$ of satellites lose all \glspl{isl}, every strategy faces cost increases of less than $10\%$. This is because the capacity of the \glspl{gsl} is the network’s bottleneck, as the following results demonstrate.}

In Fig.~\ref{fig:failplot_cost}, we present the cost evolution over six days, during which $3\%$ of the satellites experience \glspl{gsl} outages starting on the third day. Such failures commonly appear in \gls{leo} satellite networks \cite{Lai2023}. The challenge for the routing strategies is to detect this disruption and adapt by rerouting traffic through the \glspl{gsl} of unaffected satellites. On the fifth day, the network is restored to its normal state, providing an opportunity to evaluate how effectively the approaches return to regular operation. As expected, all the reference schemes experience increased costs during the third and fourth day, when network failures persist. This increase is $67.6\%$ for the bent-pipe strategy, which heavily relies on stable \glspl{gsl} and still $23.2\%$ for $k$-shortest paths. This increase is due to their inability to efficiently adapt to the reduced availability of \glspl{gsl}, leading to higher congestion. In contrast, \sysname demonstrates its resilience by quickly rerouting traffic through unaffected satellites, minimizing the impact on overall costs, which increases less than $10\%$.

The effect of rerouting data is illustrated in Fig.~\ref{fig:failplot_hops} by analyzing the average number of hops taken by data traversing the network. Notably, the majority of data is received in high-population areas near ground stations, allowing for direct transmission to the ground. As a result, the average number of hops is close to $1$ for all strategies, and for the bent-pipe strategy, it remains consistently at exactly $1$. When network failures occur, the average number of hops for \sysname and NC-\sysname increases significantly compared to the other strategies. The gradual increase seen in the figure is an artifact of the running mean. This shows that \sysname quickly detects failures and promptly reroutes traffic, effectively mitigating the impact on overall costs.
\revision{A similar effect can be observed for distributed Q-learning with the significant difference that distributed Q-learning is not able to directly return to normal operation as soon as the network recovers.}
In Fig.~\ref{fig:failplots_drop_tp}, we analyze the drop rate and throughput during the failure scenario. As expected, the drop rate increases for all reference schemes during the period of \glspl{gsl} outages. However, \sysname maintains a significantly lower drop rate compared to the other strategies. While the drop rate of $k$-shortest path increases from $5.7\%$ to $7.5\%$ and the drop rate of the bent-pipe strategy from $1.1\%$ to $3.7\%$, \sysname's average drop rate does not exceed $0.7\%$ even during the failures on the third and forth day. Note that Dijkstra's drop rate consistently exceeds $12\%$, which is why it is not included in Fig.~\ref{fig:failplot_drop}.
Correspondingly, the throughput for the reference schemes declines during the failure period, reflecting their inability to manage the rerouting efficiently. In contrast, \sysname sustains a throughput that remains close to the data generation rate, showing its resilience also under network failures.

\subsection{Parameter Optimization}

\begin{figure}
    \centering
    \includegraphics[width=\columnwidth]{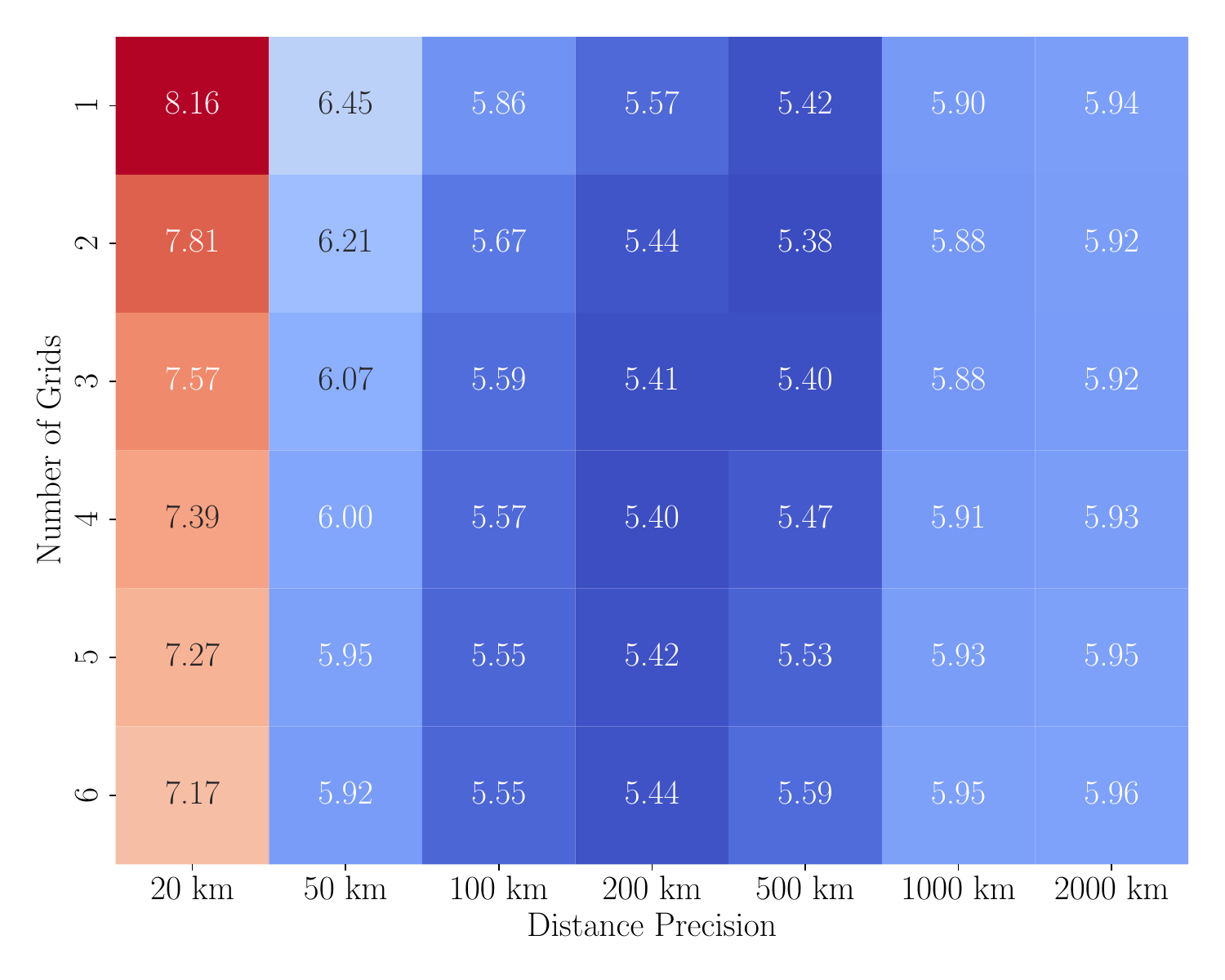}
    \caption{Average cost of \sysname for different parameters.}
    \label{fig:heatmap}
\end{figure}
\sysname uses a tile-coding mechanism described in Sec.~\ref{sec:solution}, which relies on two main parameters: The distance precision and the number of discretizations, which we explain in the following.
The continuous context space, defined by the distances to a satellite's neighbors, needs to be quantized. The precision or granularity of this quantization involves a trade-off: if the granularity is too low, \sysname fails to sufficiently distinguish between different contexts, resulting in a general solution that largely ignores the distance to neighbors. Conversely, if the granularity is too high, the number of samples per context becomes too low, making it harder for \sysname to learn effectively. The second parameter pertains to the number of overlapping partitions used in the tile-coding mechanism. This parameter also presents a trade-off: a single partition leads to overly sharp transitions between contexts, making the model overly sensitive to small changes. In contrast, too many partitions blur the distinctions between contexts, potentially masking important variations.

Both trade-offs are visualized in Fig.~\ref{fig:heatmap}. The tested granularity for the distance quantization ranges from $20$~km to $2000$~km, and the tested number of partitions ranges from $1$ to $6$. Each tile is labeled with the average cost of \sysname over a period of $7$ simulated days and is colored using a heat map scheme based on these values. Concerning the distance precision, it is clearly visible that $20$–$50$~km are too granular, resulting in a low number of samples per context, while $1000$–$2000$~km are too coarse, leading to a lack of differentiation between contexts and a generalization that overlooks variations. The lowest average cost is obtained for a precision of $500$~km.
\revision{The $500$~km quantization renders \sysname robust to errors in the estimation of the position of satellites and ground stations. Small deviations rarely change the active tile and therefore have negligible impact on the chosen actions.}
The impact of the number of partitions is smaller compared to the impact of distance precision; however, for higher granularities, it becomes increasingly important to use more partitions to avoid sharp transitions between contexts. In our evaluation, the lowest average cost was achieved using $2$ partitions. Based on these findings, we select $500$~km for the distance precision and $2$ partitions as the parameters for \sysname.

Further experiments showed that other possible contexts such as the data load at the satellite, local time of the day, UTC, or the satellite's location do not improve the performance compared to \sysname using the distance to its neighbors as per-arm context. Likewise, adding these contexts to the distance to form a larger context space does not improve the performance. Most probably, the direct influence of the distance to its neighbors on link capacity and link delay are the cause for its relevance during learning. 

%% file: sections/7a_parameters.tex
\begin{table}[htb]
\centering
\small
% \footnotesize
\renewcommand{\arraystretch}{1.6}
\begin{tabular}{|p{3.5cm}|c|c|}
\hline
\textbf{Parameter Description} & \textbf{Symbol} & \textbf{Value} \\
\hline

% RUNS
Number of repetitions & $R$ & $100$ \\
\hline

% NETWORK SIZE
Number of satellites & $N$ & $636$ \\
Number of ground stations & $M$ & $146$ \\
\hline

% SIMULATED TIME FRAME
Number of time steps & $T$ & $40320$  \\
Time step duration & $\tau$ & $15$s \\
TTL & $T_{max}$ & $200$ms \\
Simulated time frame start & - & 28.09.2023 08:26 UTC \\
\hline

% USERS
Number of users & - & $25.4$M \\
Average number of devices per person for $25.4$M users & $\mathrm{d}$ & $0.003175$ \\
Average upload traffic per device & $\nu$ & $22.98$ kbps \\
\hline

% BUFFER SIZES
Data buffer size for ground station $m$ & $Q^{\mathrm{max}}_m$ & 1GB \\
Data buffer size for sat. $n$ & $Q^{\mathrm{max}}_n$ & 50MB \\
\hline

% FIBER LINKS
Capacity of a ground station's link to internet & $C_{v}$ & $50$ Gbps \\
Delay between a ground station $m$ and the internet & $D^{\mathrm{tx}}_m$ & $1$-$5$ ms \\
\hline

% ISLS
Bandwidth for \glspl{isl} & $B_{ISL}$ & $5$ GHz \\
Aperture diameter of \gls{isl} receivers & $\varnothing$ & $10$ cm \\
\gls{isl} beam divergence angle & $\theta$ & $1.744 \cdot 10^{-5}$ Rad \\
Pointing loss factor & $L_{\text{pointing}}$ & $0.9$ \\
Noise temperature \glspl{isl} & $T_{\text{noise}}$ & $290$ K \\
\glspl{isl} transmit power & $P_{\text{tx}}$ & $0.1$ W \\
Upload scale factor & $\lambda$ & $0.08$ \\
\hline

% GSLS
Bandwidth for \glspl{gsl} & $B_{GSL}$ & $250$ MHz \\
Equivalent isotropic radiated power & EIRP & $34.6$ dbW \\
Ground station receiver antenna gain & $G_\text{rx}$ & $10.8$ db \\
\gls{gsl} carrier frequency & $f_\text{c}$ & $19$ GHz \\
Microwave background radiation temperature & $T_\text{mr}$ & $275$ K \\ 

\hline
\end{tabular}
\caption{Simulation Parameters}
\label{tab:simulation_parameters}
\end{table}

%% file: sections/8_conclusion.tex
\section{Conclusion}
\label{sec:conclusion}
In this work, we propose \sysname{}, a distributed, scalable and resilient learning approach to minimize delay and drop rate in \gls{leo} satellite networks. Considering global traffic, \sysname{} selects \glspl{isl} and \glspl{gsl} in each time slot to route user data to the internet. To address challenges like the dynamic topology of \gls{leo} satellite networks, global-scale traffic, potential network failures, and routing complexity, \sysname{} uses a contextualized \gls{mab} solution, learning link preferences based on relative distances to satellites' neighbors. It employs tile coding and the \gls{ucb} criterion for effective generalization over multiple contexts. \wds{We evaluate \sysname{} using a new simulator for global-scale simulations of stream-based data traffic.} Extensive simulations show \sysname{} outperforms reference schemes in delay, drop rate, and throughput, even under high traffic and satellite outages.

\revision{While our evaluation targets LEO, \sysname{}’s graph-based formulation, together with its context-aware link ranking and water-filling mechanism, is not restricted to LEO. Hybrid LEO–MEO–GEO and airborne operation can be realized by enlarging the node/link sets and augmenting the per-link context. In future work, we will scale our experiments to larger constellations (e.g., Starlink), incorporate alternative Quality-of-Service objectives, and integrate \gls{meo} and GEO satellites as well as airborne relays to evaluate the scalability and resilience benefits of multi-layer deployments. Additionally, a quantitative comparison of \sysname's energy consumption with fixed-path routing and centralized approaches is an interesting direction. We expect \sysname's per-satellite energy requirements to be low given its extremely small time complexity and its local-information design.}